\documentstyle[epsf]{livrev}

\RequirePackage{epsf}
\RequirePackage{longtable}

\def\fun#1#2{\lower3.6pt\vbox{\baselineskip0pt\lineskip.9pt
\ialign{$\mathsurround=0pt#1\hfil##\hfil$\crcr#2\crcr\sim\crcr}}}
\def\lap{\mathrel{\mathpalette\fun <}}
\def\gap{\mathrel{\mathpalette\fun >}}
\def\beq{\begin{equation}}
\def\eeq{\end{equation}}
\def\mh{M}
\def\msun{M_\odot}
\def\ms{m_\star}
\def\m12{m_{12}}

\begin{document}

\title{Massive Black Hole Binary Evolution}

\author{David Merritt \\
        Rochester Institute of Technology \\
        Rochester, NY USA \\
        e-mail:David.Merritt@rit.edu \\
        http://www.rit.edu/$\sim$drmsps/ \\
\and    Milo\v s Milosavljevi\' c \\
        Theoretical Astrophysics \\
	California Institute of Technology \\
	Pasadena, CA USA\\        
        e-mail:milos@tapir.caltech.edu \\
        http://www.tapir.caltech.edu/$\sim$milos/ \\
\\
%\small{(last modified: 20 September 2004)}
}

\date{}
\maketitle

\begin{abstract}
Coalescence of binary supermassive black holes (SBHs) 
would constitute the strongest sources of gravitational 
waves to be observed by LISA.
While the formation of binary SBHs during galaxy mergers
is almost inevitable, coalescence requires that the
separation between binary components first drop by a few
orders of magnitude, due presumably to interaction of
the binary with stars and gas in a galactic nucleus.
This article reviews the observational evidence for binary SBHs and
discusses how they would evolve.
No completely convincing case of a bound, binary SBH
has yet been found, 
although a handful of systems (e.g. interacting galaxies;
remnants of galaxy mergers) are now believed to contain two
SBHs at projected separations of $\lap 1$ kpc.
$N$-body studies of binary evolution in gas-free galaxies
have reached large enough particle numbers to
reproduce the slow, ``diffusive'' refilling of the binary's
loss cone that is believed to characterize binary evolution
in real galactic nuclei.
While some of the results of these simulations -- e.g. the
binary hardening rate and eccentricity evolution -- are
strongly $N$-dependent, others -- e.g. the ``damage''
inflicted by the binary on the nucleus -- are not.
Luminous early-type galaxies often exhibit depleted
cores with masses of $\sim 1-2$ times the mass of their
nuclear SBHs, consistent with the  predictions of
the binary model.
Studies of the interaction of massive binaries with
gas are still in their infancy, although much progress
is expected in the near future.
Binary coalescence has a large influence
on the spins of SBHs, even for mass ratios as extreme
as $10:1$, and evidence of spin-flips may have been
observed.
\end{abstract}


%\newpage

%===================================================================================

\section{Introduction}
\label{sec:intro}

With an ever-increasing number of secure detections, supermassive
black holes (SBHs) have evolved, in the span of a few years, 
from exotic possibilities to well-established
components of galaxies.  
While it was understood since the 1960's that the energy
sources of quasars must be gravitational \cite{robinson-65},
it was thirty years before the existence of SBHs was firmly
established, through measurements of the Keplerian
rise in the rotation velocity of stars or gas at the very
centers of galactic nuclei \cite{kr-95}.
It is now generally accepted that the formation and evolution of 
galaxies and SBHs are tightly intertwined, from the early phases of
proto-galactic formation 
\cite{silk-98},
through hierarchical build-up in CDM-like cosmogonies 
\cite{haehnelt-02},
to recent galaxy mergers 
\cite{mm-01}.

SBHs appear to be linked in fundamental ways to the dynamics of the 
stellar component in galaxies, both on large and small scales.
An astonishingly tight correlation exists between SBH mass and 
the central velocity dispersion of the stellar component,
$M_{\bullet}\sim\sigma^{\alpha}$, $\alpha\approx 4.5$ 
\cite{fm-00};
the correlation with the velocity dispersion averaged over
kiloparsec scales is weaker but still impressive
\cite{gebhardt-00,ferrarese-02}.
Similar correlations exist between
SBH mass and bulge luminosity
\cite{mclure-02,marconi-03}
and central concentration of the light
\cite{graham-01,erwin-04},
indicating that SBHs ``know'' about the
depth of the gravitational potential well in which they live.
These tight correlations probably reflect a degree of
feedback in the growth of SBHs
\cite{silk-98}.

On small scales, SBHs are embedded in stellar cusps,
parsec-scale regions where the stellar density increases
approximately as a power law with distance from the SBH
into the smallest resolvable radii 
\cite{crane-93,ferrarese-94,mf-95,gebhardt-96}.
Faint galaxies have steep nuclear density profiles, 
$\rho\sim r^{-\gamma}$, $1.5\lap\gamma\lap 2.5$, 
while bright galaxies typically have weaker cusps, 
$\gamma\lap 1$.
Steep cusps form naturally as the growth of the SBH pulls in stars
\cite{peebles-72}.
In small dense galaxies where the star-star relaxation time
is shorter than $10^{10}$ yr, steep cusps may also form
via collisional relaxation
\cite{bw-76,preto-04}.
Weak cusps may be remnants of strong cusps that were destroyed 
by binary SBHs during galaxy mergers;
in fact the structure and kinematics of galactic nuclei are now 
believed to be fossil relics of the merger process
\cite{merritt-04}.

Larger galaxies grow through the agglomeration of smaller galaxies and
protogalactic fragments.  
If more than one of the fragments contained a SBH,
the SBHs will form a bound system in the merger product
\cite{bbr-80,roos-81}.
This scenario has received considerable attention because the 
ultimate coalescence of such a pair would
generate an observable outburst of gravitational waves 
\cite{thorne-76}.
The evolution of a binary SBH
can be divided into three phases \cite{bbr-80}: 
1. As the galaxies
merge, the SBHs sink toward the center of the new galaxy 
via dynamical friction where they form a binary.  
2. The binary
 continues to decay via gravitational slingshot interactions
\cite{saslaw-74} in which
 stars on orbits intersecting the binary are ejected at velocities
 comparable to the binary's orbital velocity, while the binary's binding energy
 increases. 
3. If the binary's
 separation decreases to the point where the emission of gravitational waves becomes
 efficient at carrying away the last remaining angular momentum, 
the SBHs coalesce rapidly.  

The transition from (2) to (3) is understood to be the bottleneck of
 a SBH binary's path to coalescence, since the binary will quickly
eject all stars on intersecting orbits, thus cutting off the supply
of stars.
This is called the ``final parsec problem'' \cite{mm-03a}.
But there are other possible ways of continuing to extract 
energy and angular momentum
from a binary SBH, including accretion of gas onto the binary
system \cite{armitage-02} or refilling of the loss cone
via star-star encounters \cite{yu-02,mm-03b} or 
triaxial distortions \cite{poon-04b}.
Furthermore there is circumstantial evidence that efficient
coalescence is the norm.
The $X$-shaped radio sources \cite{dennett-02}
are probably galaxies in which
SBHs have recently coalesced, causing jet directions to flip.
The inferred production rate of the $X$-sources
is comparable to the expected merger rate of bright ellipticals,
suggesting that coalescence occurs relatively quickly following
mergers \cite{ekers-02}.
If binary SBHs failed to merge efficiently, uncoalesced binaries
would be present in many bright ellipticals, resulting in 3-
or 4-body slingshot ejections when subsequent mergers brought in
additional SBHs.
This would produce off-center SBHs, which seem to be rare or
non-existent, as well
as (perhaps) too much scatter in the $\mh-\sigma$ and $\mh-L_{bulge}$
relations \cite{haehnelt-02}.

While the final approach to coalescence of binary SBHs is not
well understood, much of their dynamical effect on the surrounding
nucleus takes place very soon after the binary forms.
The binary quickly (in less than a galactic crossing time) ejects
from the nucleus a mass in stars of order its own mass 
\cite{quinlan-96,mm-01}
significantly lowering the central density on parsec scales.
There is reasonably quantitative agreement between this model and
the observed structure of nuclei:
the ``mass deficit'' -- the stellar mass that is ``missing'' from the
centers of galaxies, assuming that they once had steep cusps like
those observed at the centers of faint ellipticals -- is of
order the black hole mass
\cite{milos-02,ravin-02,graham-04}.

While the binary SBH model is compelling, there is still not
much hard evidence in its support. 
Observationally, no bona fide binary SBH 
(i.e. gravitationally bound pair of SBHs)
has definitely been detected, although there is circumstantial
evidence (precessing radio jets; periodic outburst activity)
for SBH binaries in a number of active galaxies,
as reviewed briefly below (see \cite{komossa-03b} for a more 
complete review of this topic).
But the binary SBH model has one great advantage: 
the postulated effects
are accessible to observation, since they extend to
scales of $1-100$ pc, the distance out to which a binary SBH can
significantly influence stellar motions.
Much of the recent theoretical work in this field has been directed
toward understanding the influence of a binary SBH
on its stellar surroundings and looking for evidence
of that influence in the distribution of light 
at the centers of galaxies.

Following the definition of terms and time scales in
\S\ref{sec:prelim}, we present a brief overview of the
observational evidence for binary SBHs in \S\ref{sec:obs}.
Interaction of a binary SBH with stars
is discussed in \S\ref{sec:stars}.
The possibility of multiple SBHs in galactic nuclei,
and the implications for coalescence, are discussed
in \S\ref{sec:mult}.
\S\ref{sec:nbody} summarizes $N$-body work
on the evolution of binary SBHs, with an emphasis
on the question of binary wandering.
Observational evidence for the destruction of
nuclear density cusps is reviewed in \S\ref{sec:cusp}.
In some galaxies, the predominant source of torques
leading to decay of the binary may be gas; this
topic is reviewd in \S\ref{sec:gas}.
Finally, the influence of binary coalescence on
SBH spins is summarized in \S\ref{sec:spin}.

\section{Preliminaries}
\label{sec:prelim}

We write $m_1$ and $m_2$ for the masses of the two components of a
binary SBH, 
with $m_2\le m_1$, $q\equiv m_2/m_1$, and $m_{12}\equiv m_1+m_2$.
(We also sometimes write $M$ for the mass of the single
SBH that forms via coalescence of two SBHs of combined mass $m_{12}$.)
The semi-major axis of the binary's Keplerian orbit is $a$ and $e$ is
the orbital eccentricity.
The binary's binding energy is
\beq
\left| E\right| = {Gm_1m_2\over 2a} = {G\mu m_{12}\over 2a}
\eeq
with $\mu=m_1m_2/m_{12}$ the reduced mass.
The orbital period is
\beq
P=2\pi\left({a^3\over Gm_{12}}\right)^{1/2} = 
9.36\times 10^3\ {\rm yr} 
\left({m_{12}\over 10^8 \msun}\right)^{-1/2}
\left({a\over 1\ {\rm pc}}\right)^{3/2} .
\label{eq:period}
\eeq
The relative velocity of the two SBHs,
assuming a circular orbit, is
\beq
V_{bin} = \sqrt{Gm_{12}\over a} = 658\ {\rm km\ s}^{-1} \left({m_{12}\over 10^8 M_\odot}\right)^{1/2} \left({a\over 1\ {\rm pc}}\right)^{-1/2}.
\eeq
A binary is ``hard'' when its binding energy per unit
mass, $|E|/m_{12} = G\mu/2a$, exceeds $\sim\sigma^2$,
where $\sigma$ is the 1D velocity dispersion of the stars
in the nucleus.
The precise meaning of ``hard'' is debatable when talking
about a binary whose components are much more massive than
the surrounding stars \cite{hills-83,quinlan-96}.
For concreteness, we adopt the following definition for 
the semi-major axis of a hard binary:
\beq
a\le a_h \equiv {G\mu\over 4\sigma^2} \approx 2.7\ {\rm pc} \left(1+q\right)^{-1} 
\left({m_2\over 10^8 M_\odot}\right) \left({\sigma\over 200\ {\rm km\ s}^{-1}}\right)^{-2}.
\label{eq:ah}
\eeq

At distances $r\gg a$, stars respond to the binary as if it were
a single SBH of mass $M$.
The gravitational influence radius of a single SBH is
defined as the distance within which the force on a test mass
is dominated by the SBH, rather than by the stars.
A standard definition for $r_{\rm infl}$ is
\beq
r_{\rm infl} = {G\mh\over\sigma^2} \approx 10.8\ {\rm pc} \left({\mh\over 10^8M_\odot}\right) \left({\sigma\over 200\ {\rm km\ s}^{-1}}\right)^{-2}.
\label{eq:rh}
\eeq
Thus $r_{\rm infl} = 4(M/\mu)a_h$.
For an equal-mass binary, $r_{\rm infl}\approx 16a_h$,
and for a more typical mass ratio of $q=0.1$,
$r_{\rm infl}\approx 50 a_h$.
An alternative, and often more useful, 
definition for $r_{\rm infl}$ is
the radius at which the enclosed mass
in stars is twice the black hole mass:
\begin{equation}
M_*(r<r_{\rm infl}) = 2\mh.
\label{eq:def_rh}
\end{equation}
This definition is appropriate 
in nuclei where $\sigma$ is a strong function of radius;
it is equivalent to equation (\ref{eq:rh}) 
when the density of stars satisfies $\rho(r)=\sigma^2/2\pi Gr^2$,
the ``singular isothermal sphere,'' and when $\sigma$ is measured
well outside of $r_{\rm infl}$.
 
If the binary's semi-major axis is small enough that its 
subsequent evolution is dominated by emission of gravitational radiation,
then $\dot {a} \propto -a^{-3}$ and
coalescence takes place in a time $t_{gr}$, where \cite{peters-64}
\begin{eqnarray}
t_{gr} &=& {5\over 256 F(e)}{c^5\over G^3} {a^4\over\mu \m12^2}, \nonumber \\
\label{eq:peters0}
F(e) &=& \left(1-e^2\right)^{7/2}\left(1+{73\over 24}e^2 + 
{37\over 96} e^4\right).
\end{eqnarray}
This can be written
\begin{eqnarray}
t_{gr} &=& {5\over 16^4 F(e)}{G\mu^3 c^5\over \sigma^8 m_{12}^2}
\left({a\over a_h}\right)^4 \nonumber \\
&\approx& {3.07\times 10^{8}{\rm yr}\over F(e)} {q^3\over (1+q)^6} 
\left({m_{12}\over 10^8\msun}\right) 
\left({\sigma\over 200\ {\rm km\ s}^{-1}}\right)^{-8}\left({a\over 10^{-2} a_h}\right)^4.
\label{eq:peters}
\end{eqnarray}
This relation can be simplified by making use of the tight 
empirical correlation
between SBH mass and $\sigma$, the ``$M-\sigma$ relation.''
Of the two forms of the $M-\sigma$ relation in the literature
\cite{fm-00,gebhardt-00}, the more relevant one \cite{fm-00}
is based on the velocity dispersion measured in
an aperture centered on the SBH, which is approximately the
same quantity $\sigma$ defined above; the alternative form
\cite{gebhardt-00} defines $\sigma$ as a mean value along a
slit that extends over the entire half-light radius of the galaxy.
In terms of the central $\sigma$, the best current estimate
of the $M-\sigma$ relation is
\cite{ff-05}
\beq
\left({\mh\over 10^8\msun}\right) 
= (1.66\pm 0.24) \left({\sigma\over 200\ {\rm km\ s}^{-1}}\right)^\alpha
\label{eq:ms}
\eeq
with $\alpha= 4.86\pm 0.43$.
Combining equations (\ref{eq:peters}) and (\ref{eq:ms}) and setting
$M=\m12$, $F=1$ gives
\begin{eqnarray}
t_{gr} &\approx& 5.0\times 10^8 {\rm yr} {q^3\over (1+q)^6} 
\left({\sigma\over 200\ {\rm km\ s}^{-1}}\right)^{-3.14}\left({a\over 10^{-2}a_h}\right)^4 
\nonumber \\
&\approx& 7.1\times 10^8 {\rm yr} {q^3\over (1+q)^6} 
\left({m_{12}\over 10^8\msun }\right)^{-0.65}\left({a\over 10^{-2}a_h}\right)^4.
\label{eq:peters2}
\end{eqnarray}
Coalescence in a Hubble time ($\sim 10^{10}$ yr) requires
$a\lap 0.05 a_h$ for an equal-mass binary and
$a\lap 0.15 a_h$ for a binary with $q=0.1$.
Inducing a SBH to decay from a separation $a\approx a_h\approx 10^0$ pc
to a separation such that $t_{gr}\lap 10^{10}$ yr
is called the ``final parsec problem'' \cite{mm-03a}.
Much of the theoretical work on massive black hole binary evolution
has focussed on this problem.

%\newpage

\section{Observations of Binary Supermassive Black Holes}
\label{sec:obs}

\subsection{External Galaxies}

If a binary SBH is defined as two SBHs separated
by a distance $a\lap a_h$, 
then no completely convincing example of such a binary has yet been found.
Here we briefly review the small set of cases in which clear evidence
is seen for two, widely separated SBHs in a single system
(``dual SBHs''),
as well as the still-circumstantial evidence
for true binary SBHs.
For a more complete review of this topic, see \cite{komossa-03b}.

\subsubsection{Dual SBHs}

\begin{figure}[h]
  \def\epsfsize#1#2{0.6#1}
  \centerline{\epsfbox{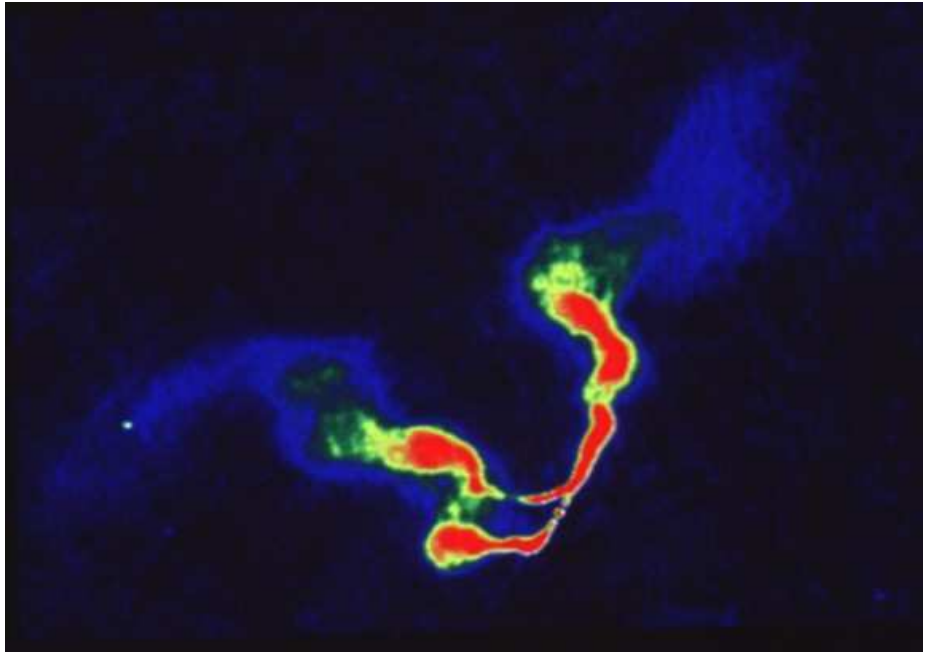}}
  \caption{\it 20 cm VLA image of the radio source 3C 75 
in the cluster of galaxies Abell 400. The image consists of two, twin-jet 
radio sources associated with each of two elliptical galaxies. 
The jets bend and appear to be interacting. The projected separation
of the radio cores is about 7 kpc. Image courtesy of NRAO/AUI and
F. N. Owen et al.}
  \label{fig:3c75}
\end{figure}

Figure~\ref{fig:3c75}
shows what was probably the first clear example of two 
SBHs in one ``system,'' in this case a pair of interacting galaxies
near the center of the galaxy cluster Abell 400.
The associated radio source 3C75 consists of a pair of twin radio lobes
originating from the radio cores of the two galaxies; the projected
separation of the cores is $\sim 7$ kpc \cite{owen-85}.
Such double-jet systems are expected to be rare given the small
fraction of giant elliptical galaxies that are associated with
luminous radio sources.

``Binary'' quasars are common but most are believed to be
chance projections or lensed images \cite{mortlock-99,kochanek-99}.
Among the binary quasars for which lensing can be ruled out,
the smallest projected separation belongs to LBQS 0103-2753 
at $z=0.85$, with an apparent spacing between centers
of $2.3$ kpc \cite{junk-01}.
However the two quasar spectra show a $\Delta z$ of $0.024$
suggesting a chance projection.

\begin{figure}[h]
  \def\epsfsize#1#2{1.3#1}
  \centerline{\epsfbox{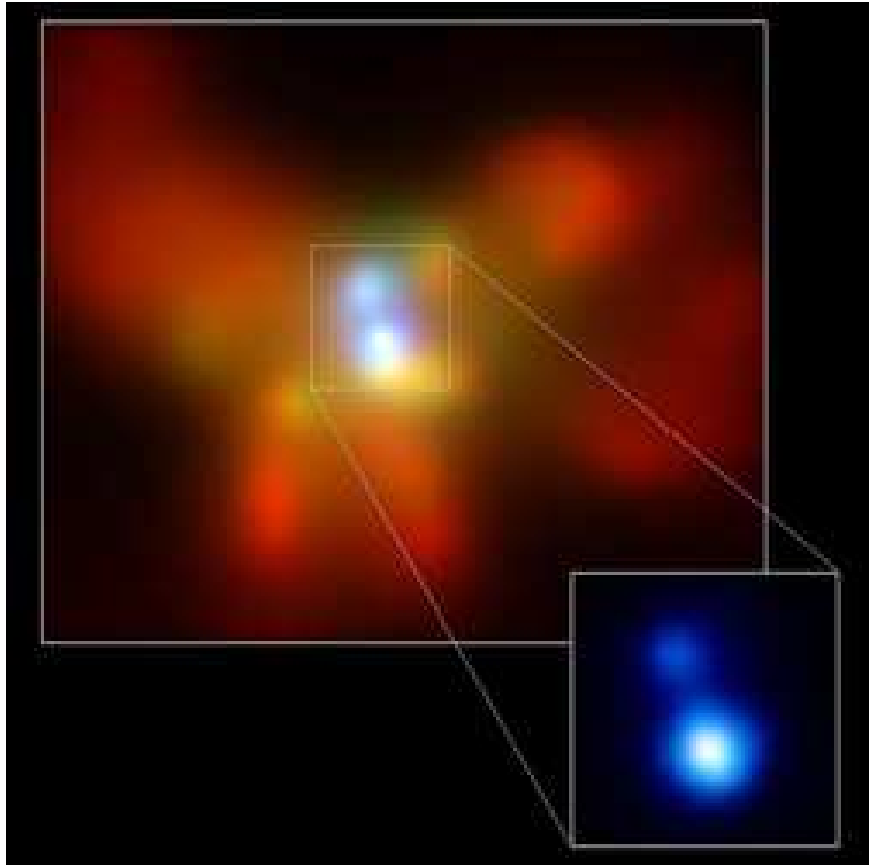}}
  \caption{\it Chandra $X$-ray image of the starburst galaxy NGC 6240, 
showing the two nuclear sources. Projected separation of the nuclei 
is about 1.4 kpc. Image courtesty of NASA/CXC/MPE/S.Komossa et al.}
  \label{fig:komossa}
\end{figure}

Galaxies in the late stages of a merger are the most plausible
sites for dual SBHs and many of these exhibit double nuclei in the
optical or infrared \cite{graham-90,carico-90}.
However few show unambiguous evidence of AGN activity in both 
nuclei, indicative of SBHs.
One clear exception is NGC 6240 (Fig.~\ref{fig:komossa}), for which both nuclei
exhibit the flat $X$-ray spectra characteristic of AGNs \cite{komossa-03}.
The projected separation is $1.4$ kpc.
Another likely case is Arp 299 \cite{ballo-04}.

Interestingly, there are no known dual SBHs with separations
below $\sim 1$ kpc, even though a $1$ kpc separation would be resolvable
to distances of several hundred Mpc.

\subsubsection{Evidence for Binary SBHs}

Many active galaxies exhibit periodic variability with periods 
of days or years, consistent with the orbital periods of true binary SBHs
having $a\lap a_h$.
Undoubtedly the clearest example is OJ 287, a ``blazar,''
i.e. an active galaxy in which the jet is believed to be orientated
nearly parallel to the line of sight, at $z=0.306$.
Optical variability of OJ 287 has been recorded since 1890 
\cite{pursimo-00,takalo-94} and has a strict period of 11.86 yr 
($\sim 9$ yr in the galaxy's rest frame); the last major outburst
was observed (on schedule) in 1994.
The outbursts are generally double-peaked with the peaks separated
by about a year; the second peak is accompanied by enhanced
radio emission.
Models to explain the periodicity usually invoke a second
SBH with $q\lap 0.1$.
In one class of model, the
variability reflects true changes in the source luminosity
due to variations in the accretion rate as the smaller SBH passes 
through the accretion disk surrounding the larger SBH
\cite{sill-88,lehto-96,valtaoja-00}.
In these models, the observed variability period is equal to the
binary orbital period, and the binary orbit is highly eccentric
($e\approx 0.7$), implying a relatively short ($\lap 10^5$ yr)
time scale for orbital decay via gravitational radiation.
The lag between primary and secondary peaks may be due to the
time required for the disturbance induced by the passage
through the accretion disk to propagate down the jet \cite{valtaoja-00}.
Alternatively, the luminosity variations may reflect 
changes in the jet direction resulting from precession
of the accretion disk, the latter induced by torques
from the second SBH \cite{katz-97}.
In this model, the binary orbital period is much 
less than 9 yr, and the secondary maxima could
be due to a ``nodding'' motion of the accretion disk
\cite{katz-97}.

Many other examples of variability in AGN at optical, radio and
even TeV energies are documented \cite{xie-03},
with periods as short as $\sim 25$ days \cite{hayashida-98}.
Indeed evidence for variability has even been claimed for the
Milky Way SBH, at radio wavelengths; the ostensible period
is 106 days \cite{zhao-01}.
However none of these examples exhibits as clear a periodicity as OJ287.

Table \ref{tab:one} gives a list of active galaxies for which
periodic variability has been claimed.

\begin{table}[h]
\begin{center}
\begin{tabular}{lll}
\hline
Source & Period (yr) & Reference \\
\hline
Mkn 421      & 23.1  & \cite{liu-97} \\
PKS 0735+178 & 14.2  & \cite{fan-97} \\
BL Lac       & 14    & \cite{fan-98} \\
ON 231       & 13.6  & \cite{liu-95} \\
OJ 287       & 11.9  & \cite{pursimo-00} \\
PKS 1510-089 & 0.92  & \cite{xie-02} \\
Sgr A$^*$    & 0.290 & \cite{zhao-01} \\
3C 345       & 10.1  & \cite{zhang-98} \\
AO 0235+16   & 5.7   & \cite{raiteri-01} \\
3C 66A       & 0.175 & \cite{lainela-99} \\
Mkn 501      & 0.065 & \cite{hayashida-98} \\
3C 273       & 0.0026& \cite{xie-99} \\
\hline
\end{tabular}
\caption{Sources with periodic variation in the nuclear emission}
\end{center}
\label{tab:one}
\end{table}

Radio lobes in active galaxies provide a fossil record of the orientation
history of the jets powering the lobes.
Many examples of sinusoidally or helically distorted jets are known,
and these observations are often interpreted via 
a binary SBH model.
The wiggles may be due to physical displacements of the
SBH emitting the jet (e.g.,~\cite{roos-93})
or to precession of the larger SBH induced by orbital motion
of the smaller SBH (e.g.,~\cite{romero-00}).
In the radio galaxy 3C 66B, the position of the radio core
shows well-defined elliptical motions with a period of just
$1.05$ yr \cite{sudou-03}, implying $t_{gr}\lap 10^3$ yr.

About a dozen radio galaxies exhibit abrupt changes in
the orientation of their radio lobes, producing
a ``winged'' or $X$-shaped morphology \cite{leahy-92}.
While originally interpreted via a precession model \cite{ekers-78},
a more likely explanation is that the SBH producing the jet
has undergone a spin flip, due perhaps to capture of
a second SBH \cite{ekers-02,zier-02}.

A number of quasars show the peaks of their broad
emission lines at very different redshifts from their
narrow emission lines, or two displaced emission line
peaks,
which might be attributed to orbital motion of the
SBHs associated with the line emitting regions
\cite{gaskell-83,gaskell-88,stockton-91,zhou-04}.
This interpretation has fallen out of favor however
since the candidate systems do not show the predicted
radial velocity variations \cite{eracleous-97}.

A number of other possibilities exist for detecting
binary SBHs, including the use of space interferometers
to measure the astrometric reflex motion of AGN photocenters
due to orbital motion of the jet-producing SBHs \cite{wehrle-03};
measurement of periodic shifts in pulsar arrival times
due to passage of gravitational waves from binary SBHs
\cite{lommen-01};
and, of course, direct detection of gravitational
waves by space-based interferometers.

\subsection{Limits on the Binarity of The Milky Way Black Hole}

\begin{figure}[h]
  \def\epsfsize#1#2{0.75#1}
  \centerline{\epsfbox{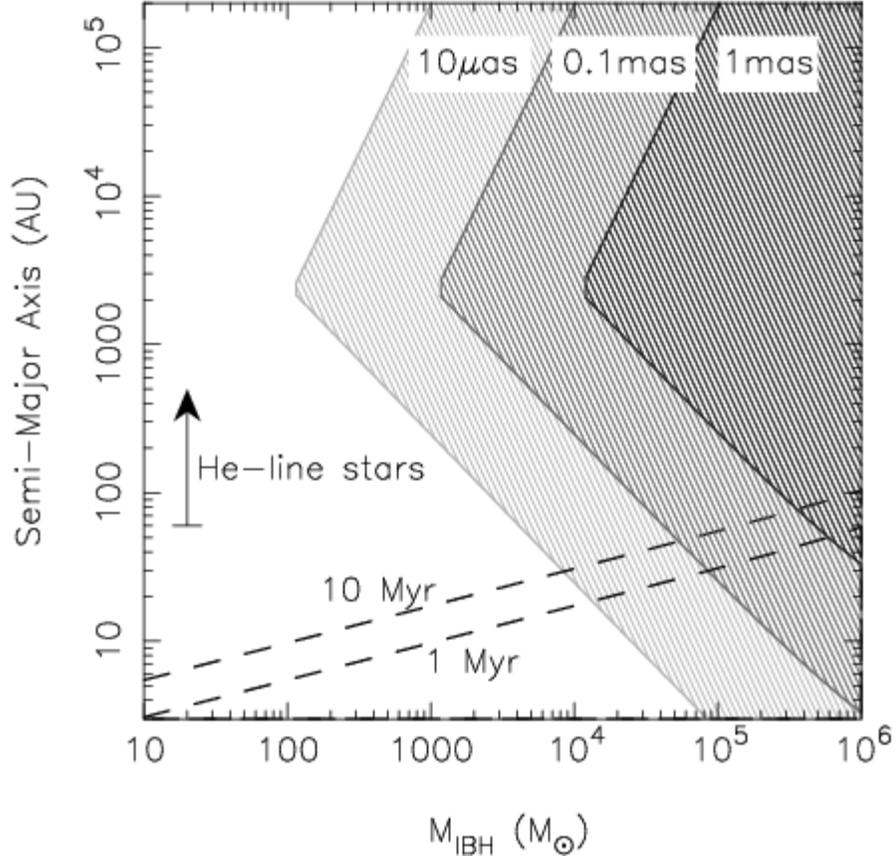}}
  \caption{\it 
   A crude illustration of the parameter space for 
a SBH-IBH binary at the Galactic center.
    Assuming a circular orbit around a SBH of $3 \times 10^6 M_{\odot}$,
a IBH with mass $M_{\rm IBH}$ 
and semi-major axis $a$ can be ruled out by measurement
of an astrometric wobble of the radio image of Sgr A$^*$.
The shaded regions show the detection thresholds
for astrometric resolutions of $0.01$, $0.1$ and 1~milliarcseconds,
respectively, assuming a monitoring period of 10 years. The dashed lines
indicate coalescence due to gravitational radiation in $10^6$ and
$10^7$~years, respectively (From \cite{hansen-03}, see also \cite{yu-03}).
  }
  \label{fig:mwbin}
\end{figure}

The likely longevity of binary SBHs motivates the question 
whether the closest and best-studied SBH at the center 
of the Milky Way galaxy is a binary.  
Monitoring of the proper motion of stars orbiting the SBH 
has led to a precise measurement of its mass, 
$M\sim4\times10^6M_\odot$ \cite{ghez-03,schoedel-03}.  
The Milky Way SBH is coincident with the compact ($<1\textrm{ AU}$) 
radio source Sagittarius (Sgr) A$^*$. 
If the Milky Way SBH were a binary, the radio source would 
probably be associated with the more massive of the two binary components.  
Limits on the masses of the components could be placed by measuring 
astrometric reflex motion of the radio source relative to distant 
quasars \cite{backer-99,reid-99}.  
Such measurements have recovered the magnitude of the solar reflex 
motion in the galaxy but have so far yielded no evidence for a binary SBH.  
The most recent upper limits on the mass of a binary companion of 
Sgr A$^*$ are $M_2\lap10^4M_\odot$ for binaries with semimajor axes 
$10^3\textrm{ AU}<a<10^5\textrm{ AU}$ \cite{reid-04}.  
This places any companion that may exist in the class of 
``intermediate-mass'' black holes (IBHs).  
The parameter space of SBH-IBH binaries at the Galactic center is 
illustrated in Figure \ref{fig:mwbin}.

IBHs have been suggested as a possible explanation for ultraluminous X-ray sources; however their existence is not widely accepted.  It has been suggested that the center of the Milky Way is a place where IBHs might naturally form via the runaway merging of massive stars in the young, dense star clusters (\cite{gurkan-04} and references therein).  Two such clusters, the Arches and the Quintuplet, are presently located in the Galactic center region. 
The segregation of massive stars to the cluster center accelerates the 
``core collapse'' in which the stellar density at the center of the cluster increases drastically. Collapse time can be shorter than the life time of the most massive stars; in this case runaway stellar coalescence ensues resulting in the formation of a supermassive star at the cluster center.  If the star survives mass loss through winds and avoids exploding as a pair-instability supernova, it collapses to form an IBH \cite{woosley-02}. Dynamical friction in the background stellar cusp of the Galactic bulge subsequently drags the IBH toward the SBH until two black holes form a hard binary.   This process might explain the puzzling presence of early-type stars \cite{genzel-03,ghez-03} deep inside the sphere of influence of the SBH at the Galactic center (\cite{hansen-03}, but see \cite{kim-04}).

\section{Interaction of Binary Black Holes with Stars}
\label{sec:stars}

\subsection{Dynamics of a Massive Binary in a Fixed Stellar Background}

\begin{figure}[h]
  \def\epsfsize#1#2{0.6#1}
  \centerline{\epsfbox{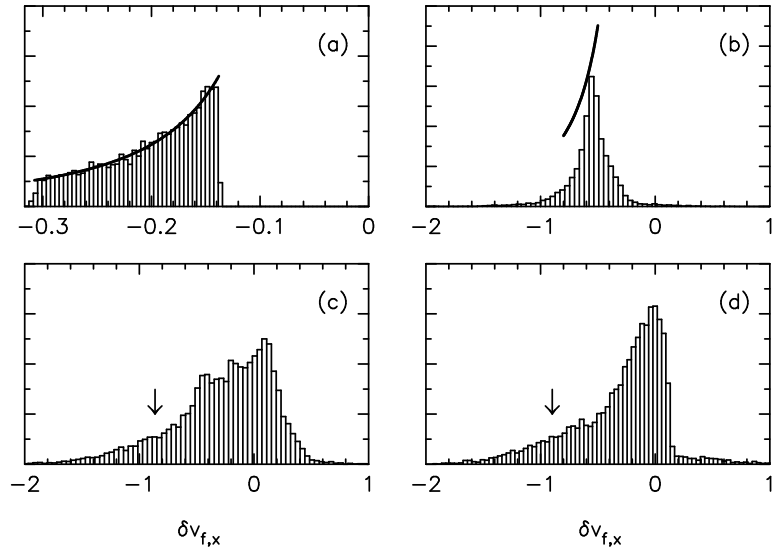}}
  \caption{\it Distribution of field star velocity changes for 
    a set of scattering experiments in which the field star's
    velocity at infinity relative to the binary was 
    ${\bf v}_{f0} = 0.5 V_{bin}\hat{\bf e}_x$.
    The binary's mass ratio was $1:1$, and the orientation of 
    the binary's orbital plane with respect to the $x$-axis was 
    varied randomly between the scattering experiments.
    Each plot represents $5\times 10^4$ scattering experiments 
    within some range of impact parameters $[p_1,p_2]$ in units of $a$. 
(a) $[6,10]$
(b) $[2,4]$
(c) $[0.6,1]$
(d) $[0.4,0.6]$.
Solid lines in (a) and (b) are the distributions corresponding to 
scattering off a point-mass perturber.
In (c) and (d), the mean of this distribution (which is very narrow) 
is indicated by the arrows.
   The gravitational slingshot is apparent in the rightward
   shift of the $\delta v$ values when $p$ is small, due to
  the randomization of ejection angles
  (from \cite{merritt-01}).}
  \label{fig:scatter}
\end{figure}

Stars passing within a distance $\sim 3a$ of the center of mass
of a hard binary undergo a complex interaction with the two
black holes, followed almost always by ejection at velocity
$\sim \sqrt{\mu/\m12} V_{bin}$, the ``gravitational slingshot''
\cite{saslaw-74}.
Each ejected star carries away energy and angular momentum,
causing the semi-major axis, eccentricity, orientation,
and center-of-mass velocity of the
binary to change and the local density of stars to drop.
If the stellar distribution is assumed fixed far from the binary
and if the contribution to the potential from the stars is ignored, 
the rate at which these changes occur
can be computed by carrying out scattering experiments
of massless stars against a binary whose 
orbital elements remain fixed during each interaction
\cite{hills-80, roos-81, hills-83, hills-92, baranov-84,
mikkola-92, quinlan-96, merritt-01, merritt-02}.
Figure \ref{fig:scatter} shows an example of field star
velocity changes in a set of scattering experiments.

Consider an encounter of a single field star of mass $\ms$
with the binary.
Long before the encounter, the field star has velocity ${\bf v}_0$ with
respect to the center of mass of the field star-binary system
and its impact parameter is $p$.
A long time after the encounter, the velocity ${\bf v}$ of the field
star attains a constant value.
Conservation of linear momentum implies that the change 
$\delta{\bf V}$ in the velocity of the binary's center of mass
is given by
\beq
\delta{\bf V} = -{\ms\over\m12} \delta{\bf v}.
\eeq
The velocity change results in a random walk of the binary's
center-of-mass momentum, as discussed in more detail below.
The energy of the field star-binary system, expressed in terms of
pre-encounter quantities, is
\begin{eqnarray}
E_0&=&{1\over 2}\ms v_0^2 + {1\over 2}\m12 V_0^2 - {Gm_1m_2\over 2a_0}
\nonumber \\
&=& {1\over 2}\ms \left(1+{\ms\over\m12}\right)v_0^2 - {Gm_1m_2\over 2a_0}
\end{eqnarray}
with ${\bf V}_0=-(\ms/\m12){\bf v}_0$ 
the initial velocity of the binary's center of mass
and $a_0$ the binary's initial semi-major axis.
After the encounter,
\beq
E= {1\over 2}\ms\left(1+{\ms\over\m12}\right)v^2 - {Gm_1m_2\over 2a}
\eeq
and $E=E_0$, so that
\begin{eqnarray}
\delta\left({1\over a}\right)&=&
{\ms(v^2-v_0^2)\over Gm_1m_2}\left(1+{\ms\over\m12}\right)
\nonumber \\
&\approx& {\ms(v^2-v_0^2)\over Gm_1m_2}.
\end{eqnarray}
\label{eq:harden}
Averaged over a distribution of field-star velocities and
directions, equation (\ref{eq:harden}) gives the binary hardening
rate $(d/dt)(1/a)$
\cite{hills-83,hills-92,mikkola-92,quinlan-96}.

The angular momentum of the field star-binary system about its center of mass, 
expressed in terms of pre-encounter quantities, is
\beq
{\bf \cal{L}}_0=\ms\left(1+{\ms\over\m12}\right){\bf \ell}_0 + 
\mu{\bf \ell}_{b0} 
\eeq
where ${\bf \ell}_0\equiv p{\bf v}_0$ and ${\bf \ell}_{b0}\equiv{\bf \cal{L}}_{b0}/\mu$
with ${\bf \cal{L}}_b$ the binary's orbital angular momentum.
Conservation of angular momentum during the encounter gives
\begin{eqnarray}
\delta{\bf \ell}_b &=& 
-{\ms\over\mu}\left(1+{\ms\over\m12}\right)\delta{\bf \ell} \nonumber \\
&\approx& -{\ms\over\mu}\delta{\bf \ell}.
\label{eq_dl}
\end{eqnarray}
Changes in $|{\bf \ell}_b|$ correspond to changes in the 
binary's orbital eccentricity $e$ via the relation 
$e^2=1-\ell_b^2/G\m12\mu^2 a$
\cite{mikkola-92,quinlan-96}.
Changes in the direction of ${\bf \ell}_b$ correspond to changes in the
orientation of the binary \cite{merritt-02}.

The results of the scattering experiments can be summarized
via a set of dimensionless coefficients $H, J, K, L, ...$
which define the mean rates of change of the parameters
characterizing the binary and the stellar background.
These coefficients are functions of the binary mass ratio,
eccentricity and hardness but are typically independent
of $a$ in the limit that the binary is very hard.
The hardening rate of the binary is given by
\begin{equation}
{d\over dt}\left({1\over a}\right) = H{G\rho\over\sigma}
\label{eq:def_H}
\end{equation}
with $\rho$ and $\sigma$ the density and 1D velocity dispersion
of stars at infinity.
The mass ejection rate is
\begin{equation}
{dM_{ej}\over d\ln(1/a)} = J m_{12}
\label{eq:def_J}
\end{equation}
with $M_{ej}$ the mass in stars that escape the binary.
The rate of change of the binary's orbital eccentricity is
\begin{equation}
{de\over d\ln(1/a)} = K.
\label{eq:def_K}
\end{equation}
The diffusion coefficient describing changes in the
binary's orientation is
\begin{equation}
{\langle\Delta\vartheta^2\rangle}= L{m_\star\over m_{12}}{G\rho a\over\sigma}
\label{eq:def_L}
\end{equation}
with $m_\star$ the stellar mass.
Brownian motion of the binary's center of mass is determined
by the coefficients $A$ and $C$ which characterize the
Chandrasekhar diffusion coefficients at low ${\bf V}$:
\begin{eqnarray}
\langle \Delta v_{\parallel}\rangle &=& -AV, \nonumber \\
\langle\Delta v_{\parallel}^2\rangle &=& {1\over 2}\langle\Delta v_{\perp}^2\rangle = C.
\end{eqnarray}
The mean square velocity of the binary's center of mass 
is $\langle V^2\rangle = C/2A$.

The binary hardening coefficient $H$ reaches a constant value
of $\sim 16$ in the limit $a\ll a_h$, with a weak dependence
on $q$ \cite{hills-83,mikkola-92,quinlan-96}.
In a fixed background, equation (\ref{eq:def_H}) therefore implies that
a hard binary hardens at a constant rate:
\begin{equation}
{1\over a(t)} - {1\over a_h} \approx 
H {G\rho \over \sigma}\left(t-t_h\right),\ \ \ \ t\ge t_h, \ \ \ \ a(t_h)=a_h.
\label{eq:decay}
\end{equation}
This is sometimes taken as the definition of a ``hard binary.''
The time to reach zero separation is 
\begin{eqnarray}
t-t_h &=& {\sigma\over HG\rho} {1\over a_h} = {2\sigma^3\over HG^2\rho\mu} =
{2q\left(1+q\right)^2\over H} {\sigma^3\over G^2\rho m_{12}}
\nonumber \\
&\approx& 5.2\times 10^5\ {\rm yr}\ q\left(1+q\right)^2
\left({\sigma\over 200\ {\rm km\ s}^{-1}}\right)^3
 \left({\rho\over 10^3 \msun\ {\rm pc}^{-3}}\right)^{-1} \nonumber \\
&\times& \left({m_{12}\over 10^8\msun}\right)^{-1}.
\label{eq:thard}
\end{eqnarray}
Orbital shrinkage would occur quite rapidly in the environment of 
a galactic nucleus if the properties of the
stellar background remained fixed.

\begin{figure}[h]
  \def\epsfsize#1#2{0.6#1}
  \centerline{\epsfbox{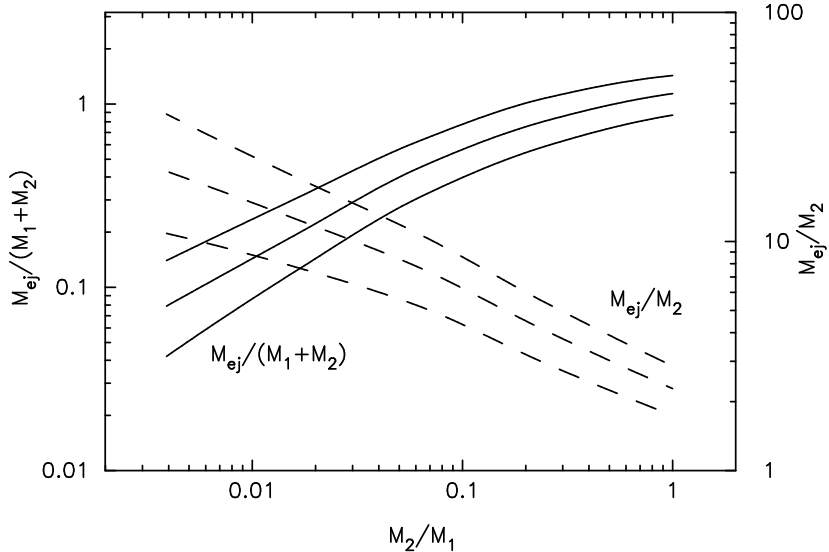}}
  \caption{\it Mass ejected by a decaying binary, 
    in units of $m_{12}=m_1+m_2$ (solid lines)
    or $m_2$ (dashed lines), calculated by an integration 
    of equation (\ref{eq:mej_int}), with the coefficient 
    $J(a)$ taken from \cite{quinlan-96}.
    Curves show mass that must be ejected in order
    for the binary to reach a separation where
    the emission of gravitational radiation
    causes coalescence on a time scale of
    $10^{10}$ yr (lower), $10^9$ yr (middle) and
    $10^8$ yr (upper).}
  \label{fig:mej}
\end{figure}

However if the binary manages to shrink to a 
separation 
at which $t_{gr}\lap 10^{10}$ yr,
the changes it induces in its stellar surroundings will be considerable.
The mass ejected by the binary in decaying from $a_h$ to $a_{gr}$
is given by the integral of equation (\ref{eq:def_J}):
\begin{equation}
M_{ej} = m_{12}\int_{a_{gr}}^{a_h} {J(a)\over a} da.
\label{eq:mej_int}
\end{equation}
Figure~\ref{fig:mej} shows $M_{ej}$ as a function of the mass ratio 
$q$ for $\sigma=200$ km s$^{-1}$ and various values
of $t_{gr}$.
The mass ejected in reaching coalescence is of order
$m_{12}$ for equal-mass binaries, and several times $m_2$
when $m_2\ll m_1$.
A SBH that grew to its current size through
a succession of mergers should therefore have displaced
a few times its own mass in stars.
If this mass came mostly from stars that were originally 
in the nucleus,
the density within $r_{\rm infl}$ would drop drastically and the
hardening would stop.
Without some way of replenishing the supply of stars
(and in the absence of other mechanisms for extracting 
angular momentum from the binary, e.g., torques from gas clouds;
cf. \S 8),
decay would stall at a separation much greater than $a_{gr}$.

In the case of extreme binary mass ratios, $m_2\ll m_1$,
the assumption that all stars passing a distance
$\sim a$ from the binary will be ejected is likely to be 
incorrect; many such stars will pass through the
binary system without being appreciably perturbed
by the smaller black hole. 
The concept of ``ejection'' in extreme binary
mass ratios may be misleading; since the typical amount of energy
transferred in  stellar pericenter passages
is small, most of the slingshot stars are not ejected from the nucleus and
remain bound to 
the larger black hole. Figure~\ref{fig:mej} should thus
be interpreted with caution in the regime $m_2\ll m_1$.
The extreme-mass-ratio regime is poorly understood 
but deserves more study
in view of the possibility that intermediate-mass black
holes may exist having masses of $\sim 10^3\msun$
\cite{fabbiano-05}.

Changes in the binary's orbital eccentricity
(eq. \ref{eq:def_K})
are potentially important because the gravity wave coalescence
time drops rapidly as $e\rightarrow 1$ (eq. \ref{eq:peters}).
For a hard binary, scattering experiments give
$K(e)\approx K_0 e (1-e^2)$, with $K_0\approx 0.5$ for an
equal-mass binary
\cite{mikkola-92,quinlan-96}.
The dependence of $K$ on $m_2/m_1$ is not well understood
and is an important topic for further study.
The implied changes in $e$ as a binary decays from $a=a_h$ to 
zero are modest, $\Delta e\lap 0.2$, for all initial eccentricities.

\subsection{Evolution in an Evolving Background}
\label{sec:gal}

The scattering experiments summarized above treat the
binary's environment as fixed and homogeneous.
In reality, the binary is embedded at the center of
an inhomogeneous and evolving galaxy, and the supply of stars 
that can interact with it is limited.

In a fixed spherical galaxy,
stars can interact with the binary only if their pericenters
lie within $\sim {\cal R}\times a$, where ${\cal R}$ is of order
unity.
Let $L_{lc}={\cal R}a\sqrt{2\left[E-\Phi({\cal R}a)\right]}\approx 
\sqrt{2Gm_{12}{\cal R}a}$,
the angular momentum of a star with pericenter ${\cal R}a$.
The ``loss cone'' is the region in phase space defined by
$L\le L_{lc}$.
The mass of stars in the loss cone is
\begin{eqnarray}
M_{lc}(a)&=&m_*\int dE \int_0^{L_{lc}} dL\ N(E,L^2) \nonumber \\
&=& m_*\int dE\int_0^{L_{lc}^2} dL^2 4\pi^2 f(E,L^2) P(E,L^2) \nonumber \\
&\approx& 8\pi^2 Gm_{12}m_* {\cal R}a\int dE f(E) P_{rad}(E).
\label{eq:def_mlc}
\end{eqnarray}
Here $P$ is the orbital period, $f$ is the number density
of stars in phase space, and $N(E,L^2)dEdL$ is the number
of stars in the integral-space volume defined by $dE$ and $dL$.
In the final line,
$f$ is assumed isotropic and $P$ has been
approximated by the period of a radial orbit of energy $E$.
An upper limit to the mass that is available to interact with 
the binary 
is $\sim M_{lc}(a_h)$, the mass within the loss cone when the
binary first becomes hard; 
this is an upper limit since some stars that are initially within
the loss cone will ``fall out'' as the binary shrinks.
Assuming a singular isothermal sphere for the stellar distribution,
$\rho\propto r^{-2}$, and taking the lower limit of the energy
integral to be $\Phi(a_h)$, 
equation (\ref{eq:def_mlc}) implies
\begin{equation}
M_{lc}(a_h) \approx 3{\cal R}\mu.
\label{eq:mlc}
\end{equation}
We can compute the change in $a$ that would result if
the binary interacted with this entire mass, by using
the fact the mean energy change of a star interacting 
with a hard binary is $\sim 3G\mu/2a$ \cite{quinlan-96}.
Equating the energy carried away by stars with
the change in the binary's binding energy gives
\begin{equation}
{3\over 2}{G\mu\over a} dM \approx {Gm_1m_2\over 2}d\left({1\over a}\right)
\label{eq:dm1}
\end{equation}
or
\begin{equation}
\ln\left({a_h\over a}\right) \approx 3{\Delta M\over m_{12}} \approx
{9{\cal R}\mu\over m_{12}}\approx 9{\cal R} {q\over (1+q)^2}
\label{eq:dm2}
\end{equation}
if $\Delta M$ is equated with $M_{lc}$.
Only for very low mass ratios ($q\lap 10^{-3}$) 
is this decay factor large enough to give $t_{gr}<10^{10}$ yr
(eq. \ref{eq:peters}), but the time required for 
such a small black hole
to reach the nucleus is likely to exceed a Hubble time
\cite{merritt-00}.
Hence even under the most favorable assumptions, the
binary would not be able to interact with enough mass
to reach gravity-wave coalescence.

But the situation is even worse than this, since
not all of the mass in the loss cone will find its
way into the binary.
The time scale for the binary to shrink is comparable with
stellar orbital periods, and some of the stars with $r_{peri}\approx a_h$
will only reach the binary after $a$ has fallen below $\sim a_h$.
We can account for the changing size of the loss cone by writing
\begin{eqnarray}
{dM\over dt} &=& \int_{E_0(t)}^{\infty} {1\over P(E)} {d M_{lc}\over dE} dE 
\nonumber \\ 
&=& 8\pi^2 Gm_{12}m_* {\cal R}a(t)\int_{E_0(t)}^{\infty} f_i(E)dE,
\label{eq:dm3}
\end{eqnarray}
where $M(t)$ is the mass in stars interacting with the binary and
$f_i(E)$ is the initial distribution function; setting
$P(E_0)=t$ reflects the fact that stars on
orbits with periods less than $t$ have already interacted with the binary
and been ejected.
Combining equations (\ref{eq:dm1}) and (\ref{eq:dm3}),
\begin{equation}
{d\over dt}\left({1\over a}\right) \approx 24\pi^2{\cal R}G m_* \int_{E_0(t)}^{\infty} f_i(E) dE.
\label{eq:shrink}
\end{equation}
Solutions to equation (\ref{eq:shrink}) show that a binary in
a singular isothermal sphere galaxy stalls at $a_h/a\approx 2.5$ for
$m_2= m_1$, compared with $a_h/a\approx 10$ if the full loss cone
were depleted (equation~\ref{eq:dm2}).
In galaxies with shallower central cusps, decay of the binary would stall
at even greater separations.

\subsection{Collisional Loss-Cone Replenishment}
\label{sec:collision}

A binary black hole depletes its loss cone
very quickly since
stars within the loss cone need only a few close
encounters with the binary to be ejected.
Whether the binary can continue to exchange energy with stars
depends on the efficiency with which stars are re-supplied to the loss cone.

The most commonly invoked mechanism for loss cone re-filling
is two-body scattering of stars.
A small angular momentum perturbation, for instance
from a passing star,
can deflect a star with $L\gap L_{lc}$
into the loss cone.
This process has been studied in detail in the context
of scattering of stars into the tidal 
disruption sphere of a single black hole \cite{FR-76,LS-77,CK-78}.
The basic equations are similar in the case of scattering
into a binary SBH, except that the critical angular
momentum increases by a factor $\sim \sqrt{a/r_t}$,
where $r_t$ is the tidal disruption radius.
(Other differences are discussed below.)
If the binary parameters are assumed fixed,
a steady-state flow of stars into the loss cone will be 
achieved on roughly a two-body relaxation time scale $T_R$, 
and the distribution function near $L_{lc}$ will have
the form
\begin{equation}
f(E,L)\approx {1\over\ln(1/R_{lc})} \overline{f}(E) 
\ln\left({R\over R_{lc}}\right),
\label{eq:ls}
\end{equation}
where $R$ is a scaled angular momentum variable,
$R\equiv L^2/L_c^2(E)$, $L_c(E)$ is the angular momentum
of a circular orbit of energy $E$, 
and $\overline{f}$ is the distribution function
far from the loss cone, assumed to be isotropic.
The mass flow into the central object is 
$m_*\int{\cal F}(E) dE$, where
\begin{equation}
{\cal F}(E) dE = 4\pi^2 L_c^2(E) \left\{\oint {dr\over v_r} \lim_{R\rightarrow 0} {\langle(\Delta R)^2\rangle\over 2R}\right\} {\overline{f}\over \ln(1/R_{lc})} dE.
\label{eq:relax}
\end{equation}
The quantity in brackets is the orbit-averaged diffusion coefficient
in $R$.

A crude estimate of the collisional re-supply rate is given by
\beq
\dot{M_\star} \approx {M_\star(r<r_{\rm crit})\over T_R(r_{\rm crit})}
\eeq
\cite{FR-76},
where $M_\star(r)$ is the mass in stars within radius $r$ and
$r_{\rm crit}$ is the critical radius at which stars scatter
into the loss cone in a single orbital period; beyond $r_{\rm crit}$,
the diffusion rate drops rapidly with radius.
Estimates based on simple galaxy models give 
$r_{\rm crit}\approx 10-100 a$.
To get an idea of the scattering rate, 
we consider the nucleus of the Milky Way.
Equation (\ref{eq:ah}) gives $a_h\approx 0.32(1+q)^{-1}$ pc.
Assuming $a=a_h$, $q=0.1$ and $r_{\rm crit}\approx 30 a_h$ gives
$r_{\rm crit}\approx 1.1$ pc.
The mass within this radius is $\sim 3\times 10^6\msun$
\cite{genzel-03} and the relaxation time at this radius, assuming stars
of a solar mass, is $\sim 2\times 10^9$ yr.
The scattered mass over $10^{10}$ yr is then $\sim 10^7\msun$.
This is comparable to the mass of the Milky
Way SMBH, $M\approx 3\times 10^6\msun$ \cite{schoedel-03},
but the scattering rate would drop as the binary shrinks,
suggesting that scattered stars would contribute only
modestly to refilling of the loss cone.
In more massive galaxies, the nuclear density is lower,
relaxation times are longer, and collisional refilling
would be even less important.
A more detailed calculation of the collisional refilling
rates in real galaxies \cite{yu-02} concludes that few if
any binary SBHs could reach coalescence via this mechanism. The author
took the presently observed luminosity profiles of
galaxies as initial conditions. The stellar
density profiles must have been steeper before the binary
SBH formed \cite{mm-01}, leading to substantially
more rapid decay early in the life of the binary.

\begin{figure}[h]
  \def\epsfsize#1#2{0.75#1}
  \centerline{\epsfbox{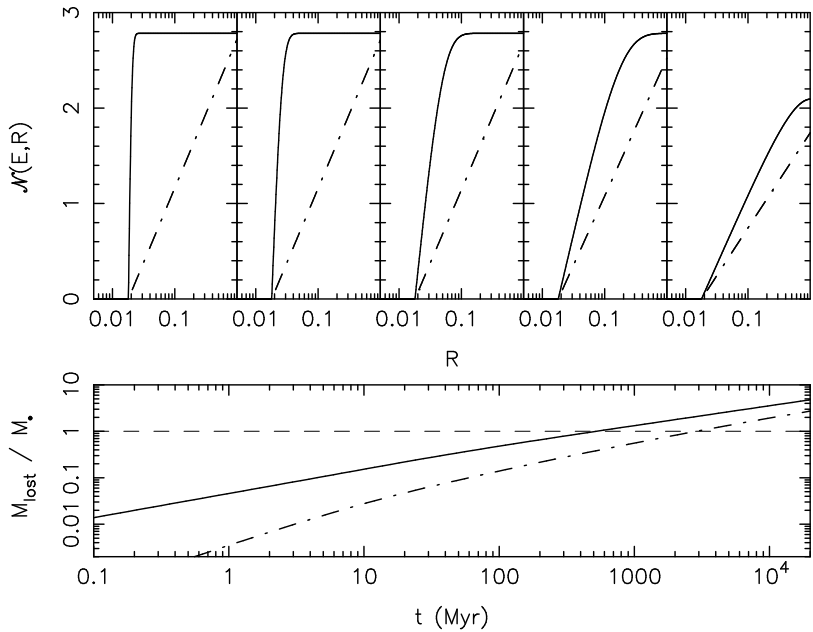}}
  \caption{\it 
(a) Slices of the density $N(E,R,t)$ at one,
  arbitrary $E$, recorded, from left to right, at $10^0$, $10^1$, $10^2$,
  $10^3$, and $10^4$ Myr (solid
  curve). Initially, $N(E,R,t)=0$ for $R\leq R_{lc}$ and
  $N(E,R,t)=\textrm{const}$ for $R>R_{lc}$.  We also show the
  equilibrium solution of equation (\ref{eq:ls}) (dot-dashed
  curve).  
(b) The total number of stars consumed by the
  loss cone as a function of time (solid curve).  
  The scale has been set to galaxy M32
  with
  initial separation between the MBHs of
  0.1 pc.
(From \cite{mm-03b})
}
  \label{fig:slice}
\end{figure}

Another criticism of standard loss cone theory is its assumption
of a quasi-steady-state
distribution of stars in phase space near $L_{lc}$ \cite{mm-03b}.
This assumption is appropriate at the center of a globular cluster,
where relaxation times are much shorter than the age of the universe,
but is less appropriate for a galactic nucleus,
where relaxation times almost always greatly exceed a Hubble time 
\cite{faber-97}.
(The exceptions are the nuclei of small dense systems like the
bulge of the Milky Way.)
The distribution function $f(E,L)$ immediately following the
formation of a hard binary is approximately a step function,
\begin{equation}
f(E,L) \approx \cases {\overline{f}(E),& $L>L_{lc}$\cr 0,&$L<L_{lc}$,\cr}
\end{equation}
much steeper than the $\sim\ln L$ dependence in a collisonally
relaxed nucleus (equation \ref{eq:ls}).
Since the transport rate in phase space is proportional
to the gradient of $f$ with respect to $L$, steep gradients imply
an enhanced flux into the loss cone.
Figure \ref{fig:slice} shows the evolution of $N(E,R)$
at a single $E$ assuming that the loss cone is empty initially
within some $R_{lc}\equiv L_{lc}/L_c(E)$ and that $N(E,R,t=0)$ is a
constant function of $R$ outside of $R_{lc}$.
(The loss cone boundary is assumed static; in reality it would
shrink with the binary.)
Also shown is the collisionally-relaxed solution of equation (\ref{eq:ls}).
The phase-space gradients decay rapidly at first and then
more gradually as they approach the steady-state solution.
The total mass consumed by the binary, shown in the lower panel
of Fig. \ref{fig:slice}, is substantially greater than would
be computed from the steady-state theory,
implying greater cusp destruction and more rapid decay of the binary.
This time-dependent
loss cone refilling might be particularly effective in a nucleus
that that continues to experience mergers or accretion events, in such
a way that the loss cone repeatedly returns to an unrelaxed state
with its associated steep gradients.

There are other differences between loss cones around single
and binary SBHs.
A star that interacts with a massive binary generally 
remains inside the galaxy and 
is available for further interactions.
In principle, a single star can interact many times with the binary
before being ejected from the galaxy or falling outside the loss
cone; each interaction takes additional energy from the binary
and hastens its decay.
Consider a simple model in which a group of $N$ stars in a spherical
galaxy interact with the binary and receive a mean energy increment 
of $\langle\Delta E\rangle$.
Let the original energy of the stars be $E_0$.
Averaged over a single orbital period $P(E)$, the binary hardens
at a rate
\begin{equation}
{d\over dt}\left({Gm_1m_2\over 2a}\right) = m_* {N\langle\Delta E\rangle\over P(E)}.
\end{equation}
In subsequent interactions, the number of stars that remain inside
the loss cone scales as $L_{lc}^2\propto a$ while the ejection energy
scales as $\sim a^{-1}$.
Hence $N\langle\Delta E\rangle\propto a^1a^{-1}\propto a^0$.
Assuming the singular isothermal sphere potential for the galaxy,
one finds
\begin{equation}
{a_h\over a(t)} \approx 1 + {\mu\over m_{12}}\ln\left[ 1 + {m_*N\langle\Delta E\rangle\over 2\mu\sigma^2}{t-t_h\over P(E_0)}\right]
\end{equation}
\cite{mm-03b}.
Hence the binary's binding energy increases as the logarithm of the time,
even after all the stars in the loss cone have interacted at least 
once with the binary.
Re-ejection would occur differently in nonspherical galaxies
where angular momentum is not conserved and ejected stars could miss
the binary on their second passage. However there will generally exist
a subset of orbits defined by a maximum pericenter distance $\lap a$
and stars scattered onto such orbits can continue to interact with 
the binary.

As these arguments suggest, the long-term evolution of a binary
SBH due to interactions with stars may be very different
in different environments.
(We stress that the presence of gas may substantially alter
this picture; cf. \S 8.)
There are three characteristic regimes 
(we stress that the presence of gas 
 may substantially alter this picture; cf. section 8)
\cite{mm-03b}.

1. {\it Collisional.} The relaxation time $T_R$ is shorter than the
lifetime of the system and the phase-space gradients at the edge of
the loss cone are given by steady-state solutions to the Fokker-Planck
equation.
The densest galactic nuclei may be in this regime.
Resupply of the loss cone takes place on the time scale
associated with scattering of stars onto eccentric orbits.
The decay
time of a binary SBH scales as $|a/\dot a|\sim m_*^{-1}\sim N$
with $N$ the number of stars.
In the densest galactic nuclei, collisional loss cone
refilling may just be able to drive a binary SBH to coalescence
in a Hubble time.
For sufficiently small $T_R$, scattering refills the
loss cone in less than an orbital period (``full loss cone'')
and the decay follow $a^{-1}\sim t$.
$N$-body studies are typically in this regime, 
as discussed below.

2. {\it Collisionless.} The relaxation time is longer than
the system lifetime and gravitational encounters between
stars can be ignored.
The low-density nuclei of bright elliptical galaxies
are in this regime.
The binary SBH quickly interacts with stars whose pericenters
lie within its sphere of influence; in a low-density 
(spherical or axisymmetric) nucleus,
the associated mass is less than that of the binary and 
the decay tends to stall at a separation too large for gravitational wave
emission to be effective.
However evolution can continue due to re-ejection of stars
that lie within the binary's loss cone but have not yet escaped from
the system.
In the spherical geometry, re-ejection implies $|a/\dot a|\sim 
(1+t/t_0)/a$, leading to a logarithmic dependence of binary
hardness on time.
Re-ejection in galactic nuclei may contribute a factor of
$\sim$ a few to the change in $a$ over a Hubble time.

3. {\it Intermediate.} The relaxation time is of order the age of the system
or somewhat longer.
While gravitational encounters contribute to the re-population of the
loss cone, not enough time has elapsed for the phase space
distribution to have reached a collisional steady state.
Most galactic nuclei are probably in this regime.
The flux of stars into the loss cone can be substantially higher
than predicted by the steady-state theory, due to strong
gradients in the phase space density near the loss cone boundary
produced when the binary SBH initially formed.
This transitory enhancement would be most important in a nucleus that continues
to experience mergers or infall, in such a way that the loss
cone repeatedly returns to an unrelaxed state with its associated 
steep gradients.

Table 1 summarizes the different regimes.
The evolution of a real binary SBH may reflect a combination
of these and other mechanisms, such as interaction with gas.
There is a close parallel between the final parsec problem
and the problem of quasar fueling: both requre that of order 
$10^8M_\odot$ be supplied to the inner parsec of a galaxy
in a time shorter than the age of the universe.
Nature clearly accomplishes this in the case of quasars,
probably through gas flows driven by torques from stellar bars.
The same inflow of gas could contribute to the decay of a 
binary SBH in a number of ways: by leading to the renewed formation
of stars which subsequently interact with the binary;
by inducing torques which extract angular momentum from the binary;
through accretion, increasing the masses of one or both
of the SBHs and reducing their separation; etc.

\begin{table}
\caption{Physical Regimes for Long-Term Decay of Massive Black Hole Binaries}
\begin{tabular}{ll} \hline \hline
\emph{Form of Decay} & \emph{Regime} \\ 
\hline
$a^{-1} \propto \textrm{const}$ & Collisionless \\
$\phantom{a^{-12}} \propto t/N$ & Collisional (diffusion) \\
$\phantom{a^{-12}} \propto t + \textrm{const}$ & Collisional (full loss cone) \\
$\phantom{a^{-12}} \propto \ln(1 + t/t_0) + \textrm{const}$ & Re-ejection \\
\hline
\end{tabular}
\label{tab:regimes}
\end{table}

\subsection{Non-axisymmetric nuclei}
\label{sec:nonaxi}

The estimates made above were based on spherical models of nuclei.
The total number of stars in a full loss cone can be much larger if the nucleus is flattened and axisymmetric \cite{mt-99}, when only one component of the angular momentum is conserved.  In very flattened nuclei (with ellipticities $\epsilon\sim 0.5$), single emptying of an initially full loss cone can in some cases be sufficient to drive the binary to coalescence \cite{yu-02}.  
%Things do not change greatly when nuclei are flattened and axisymmetric,
%since orbits in axisymmetric potentials still conserve one 
%component of the angular momentum.
However loss cone dynamics can be qualitatively different in 
non-axisymmetric (triaxial or bar-like) potentials,
since a much greater number of stars may be on 
``centrophilic'' -- box or chaotic -- orbits 
which take them arbitrarily near to the SBH(s)
\cite{norman-83,gerhard-85,ms-93,valluri-98,yu-02}.
Stars on centrophilic orbits of energy $E$ experience pericenter passages
with $r_{\rm peri}<d$ at a rate $\sim A(E)d$
\cite{poon-04b}.
If the fraction of stars on such orbits is appreciable,
the supply of stars into the binary's loss cone will
remain essentially constant, 
even in
the absence of collisional loss-cone refilling.
Such models
need to be taken seriously given recent
demonstrations \cite{poon-02,holley-02,poon-04} that
galaxies can remain stably triaxial even
when composed largely of centrophilic orbits.
Furthermore imaging of galaxy centers on
parsec scales reveals a wealth of features in the
stellar distribution that are not consistent with axisymmetry,
including bars, nuclear spirals, and other misaligned features
\cite{wozniak-95,peng-02,erwin-02}.

The total rate at which stars pass within a distance ${\cal R}a$ of
the massive binary is 
\begin{equation}
{dM_\star\over dt} \approx {\cal R}a\int A(E)M_c(E)dE
\end{equation}
where $M_c(E)dE$ is the mass on centrophilic orbits in the
energy range $E$ to $E+dE$.
In a nucleus with $\rho\sim r^{-2}$,
the implied feeding rate into a radius $r_{\rm infl}$ is roughly
\begin{eqnarray}
\dot M_\star &\approx& \overline{f_c} {\sigma^3\over G} \\
&\approx& 
 2500\msun\ {\rm yr}^{-1}\overline{f_c} \left({\sigma\over 200\ {\rm km\ s}^3}\right)^3,
\end{eqnarray}
where $\overline{f_c}$ is the fraction of stars on
centrophilic orbits. 
If this rate were maintained, the binary would interact with
its own mass in stars in a time of only $\sim 10^5$ yr,
similar to the decay time estimated above
(eq. (\ref{eq:thard}) for a binary in a fixed background.
In fact, the feeding rate would decline with time as 
the centrophilic orbits were depleted.
Solving the coupled set of equations for $a(t)$ and $M_c(t)$,
one finds that at late times, the binary separation
in a $\rho\propto r^{-2}$ nucleus varies as \cite{poon-04}
\beq
{a_h\over a} \approx 3\times 10^4 \overline{f_c}^2 \left({\sigma\over 200\ \mathrm{km}\ \mathrm{s}^{-1}}\right)^3 \left({\m12\over 10^8\msun}\right)^{-1} \left({t\over 10^{10} \mathrm{yr}}\right). 
\label{eq:aoft}
\eeq
Comparison with Table 1 shows that this is the same time dependence
as for the ``full loss cone'' regime of spherical nuclei.
Placing just a few percent of a galaxy's mass on centrophilic orbits
is sufficient to overcome the final parsec problem
and induce coalesence,
if the stellar density profile is steep and if the chaotic orbits 
are present at all energies.
This example is highly idealized, but shows that departures from
axial symmetry in galactic nuclei can greatly affect the 
rate of decay of a binary SBH.

\section{Multiple Black Hole Systems}
\label{sec:mult}

If binary decay stalls, 
an uncoalesced binary may be present 
in a nucleus when a third SBH, or a second binary,
is deposited there following a subsequent merger.
The multiple SBH system that forms will engage in its 
own gravitational slingshot interactions, 
eventually ejecting one or more of the
SBHs from the nucleus and possibly from the galaxy
and transferring energy to the stellar fluid.

If the infalling SBH is less massive than either
of the components of the pre-existing binary,
$m_3<(m_1,m_2)$,
the ultimate outcome is likely to be ejection of the 
smaller SBH and recoil of the binary, with the binary
eventually returning to the galaxy center.
The lighter SBH is ejected with a velocity roughly $1/3$ the relative
orbital velocity of the binary \cite{saslaw-74,hut-92},
and the binary recoils with a speed that is lower
by $m_3/(m_1+m_2)$.
Each close interaction of the smaller SBH with the binary
increases the latter's binding energy by 
$\langle E/E\rangle \approx 0.4 m_3/(m_1+m_2)$
\cite{hills-80}.
If $m_3>m_1$ or $m_3>m_2$, there will most often
be an exchange interaction, with the lightest SBH
ejected and the two most massive SBHs forming a
binary; further interactions then proceed as in 
the case $m_3<(m_1,m_2)$.

During the three-body interactions, both the semi-major
axis and eccentricity of the dominant binary change
stochastically.
Since the rate of gravity wave emission is a strong function
of both parameters ($\dot E\propto a^{-4}(1-e^2)^{-7/2}$),
the timescale for coalescence can be enormously shortened.
This may be the most promising way to
coalesce SBH binaries in the low-density nuclei of massive galaxies,
where stalling of the dominant binary is likely.

This process has been extensively modelled 
using the PN2.5 approximation to represent gravitational wave
losses \cite{peters-63} and assuming a fixed
potential for the galaxy 
\cite{valtaoja-89,mv-90,valtonen-94}.
In these studies, 
there was no attempt to follow the pre-merger evolution
of the galaxies or the interaction of the binary SBHs with
stars.
In two short non-technical contributions (submissions for
the IEEE Gordon Bell prizes in 2001 and 2002), 
J. Makino and collaborators
mention two $N$-body simulations of triple
SBH systems at the centers of galaxies using 
the GRAPE-6, and (apparently) a modified version of NBODY1.
Relativistic energy losses were neglected and the SBH particles
all had the same mass.
Plots of the time evolution of the orbital parameters
of the dominant binary show strong and chaotic
eccentricity evolution,
with values as high as $0.997$ reached for short periods.
Such a binary would lose energy by gravity wave emission 
very rapidly, by a factor $\sim 10^8$ at the time of
peak $e$ compared with
a circular-orbit binary with the same semi-major axis.

In a wide, hierarchical triple, $m_3\ll (m_1,m_2)$,
the eccentricity of the dominant binary
oscillates through a maximum value of $\sim \sqrt{1-5\cos^2i/3}$,
$|\cos i|<\sqrt{3/5}$,
with $i$ the mutual inclination angle \cite{kozai-62}.
One study \cite{blaes-02}
estimates that the coalescence time of the dominant binary
in hierarchical triples can be reduced by factors of
$\sim 10$ via the Kozai mechanism.

If the binary SBH is hard when the third SBH falls in,
the ejected SBH can gain enough velocity to escape the galaxy.
If the three masses are comparable, even the binary can 
be kicked up to escape velocity.
One study \cite{volonteri-03} estimates (based on a very
simplified model of the interactions) that the recoil
velocity of the smallest SBH is larger than galactic
escape velocities in
99\% of encounters and that the binary escapes in 8\% of
encounters.
Thus a signficant fraction of nuclei could be left with
no SBH, with an offset SBH, or with a SBH whose mass is
lower than expected based on the $\mh-\sigma$ or $\mh-L_{bulge}$
relations.

There is a need for simulations of multiple-SBH systems
that include both gravitational loss terms, accurate (regularized)
interactions between the SBH particles, and the
interactions of SBH particles with stars.

\section{$N$-Body Studies of Binary Black Hole Evolution}
\label{sec:nbody}

The interaction of a massive binary with point perturbers
at the center of a galaxy is a straightforward 
problem for $N$-body simulation.
In principle, $N$-body studies can reveal both the long-term evolution
of the binary, as well as the effect of the binary on its
stellar surroundings.
The latter can be compared with observed nuclear density
profiles as a test of the theory (\S\ref{sec:cusp}).

Unfortunately, unless great care is taken, $N$-body studies
are likely to give misleading results.
This follows from the result (Section 4.3) that time scales for 
two-body scattering of stars into the binary's loss cone
are of order $10^{10}$ yr or somewhat longer in real galaxies.
In $N$-body simulations, relaxation times are shorter 
by factors of $\sim N/10^{11}$ than in real galaxies, 
hence the long-term evolution
of the binary is likely to be dominated by spurious
loss cone refilling,
wandering of the binary,
and other noise-driven effects.

$N$-body studies are most useful at characterizing the
early stages of binary formation and decay, or 
simulating the disruptive tidal effects of a SBH on the nucleus
of an infalling galaxy.
Due to algorithmic limitations -- primarily the difficulty
of integrating galaxy models with high central concentrations --
 most such studies 
\cite{ebisuzaki-91,governato-94,makino-96,makino-97,nakano-99a,
nakano-99b,hemsendorf-02,chatterjee-03,makino-04}
have been based on galaxy models with unrealistically
large cores.

\begin{figure}[h]
  \def\epsfsize#1#2{0.6#1}
  \centerline{\epsfbox{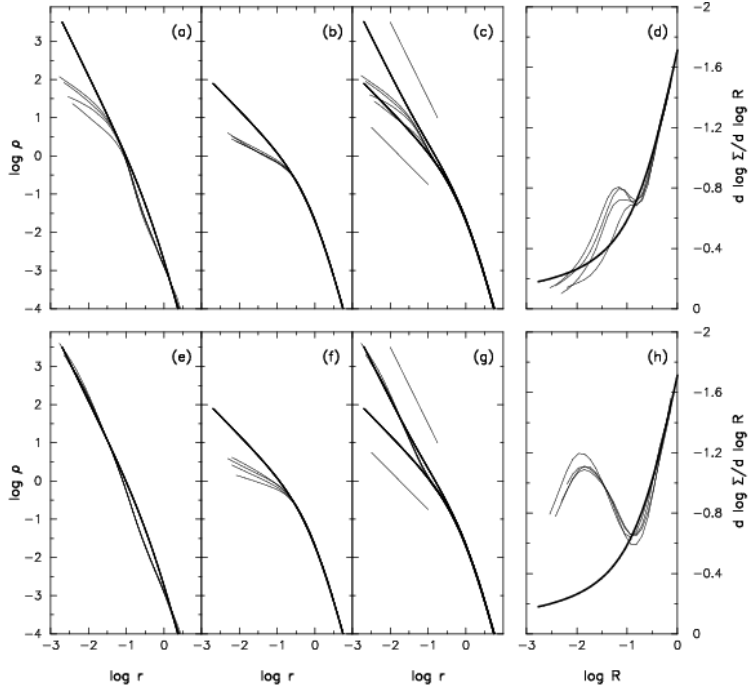}}
  \caption{
\it Final density profiles from a set of 10:1 merger simulations in
which each galaxy contained a black hole (a-d)
and in which neither galaxy contained a black hole (e-g) \cite{cruz-01}.
The four thin curves in each frame correspond to four different
pre-merger orbits.
(a), (e) Space density of stars initially associated with the secondary galaxy;
thick curves are the initial density profile.
(b), (f) Space density of stars initially associated with the primary galaxy;
thick curves are the initial density profile.
(c), (g) Space density of all stars.
Lower thick curves are the initial density profile of the primary galaxy,
and upper thick curves are the superposition of the initial density profiles
of the primary and secondary galaxies.
Lines of logarithmic slope $-1$ and $-2$ are also shown.
(d), (h) Logarithmic slope of the surface density profiles of the merger
remnants.
Thick curves correspond to the initial primary galaxy.
       }
  \label{fig:cruz}
\end{figure}

Figure~\ref{fig:cruz} is from the first \cite{cruz-01} $N$-body simulation
of galaxy mergers in which the pre-merger galaxies contained
power-law nuclear cusps as well as massive particles representing the
SBHs.
These simulations were run using {\tt GADGET} \cite{springel-01}, 
a tree code with
inter-particle softening, and were not able to accurately
follow the formation and decay of the massive binary.
The SBH in the larger galaxy was found to tidally disrupt
the steep cusp in the infalling galaxy, producing a remnant
with only slightly higher central density than that of the
giant galaxy initially.
This result helps to explain the absence of dense cusps in
bright galaxies \cite{forbes-95},
and suggests that the central structure
of galaxies can only be understood by taking into account 
the destructive influence of SBHs on the stellar distribution during mergers.
Additional results, using a similar $N$-body code and a variety of mass
ratios for the merging galaxies, were reported in \cite{mmvj-02}.

Had these simulations been extended to longer times using a more accurate
$N$-body code, the massive binary would have ejected stars
via the gravitational slingshot and lowered the
central density still more.
This was first demonstrated \cite{mm-01} in an $N$-body study that used
a tree code for the early stages of the merger,
and {\tt NBODY6}, a high-precision, direct-summation code \cite{aarseth-99},
for the later stages, when the binary separation fell below
the tree code's softening length.
The pre-merger galaxies had steep, 
$\rho\sim r^{-2}$ density cusps and
the mass ratio was 1:1.
These simulations were continued until 
the binary separation had decayed 
by a factor of $\sim 10$ below $a_h$.
The initially steep nuclear cusps were converted to
shallower, $\rho\sim r^{-1}$ profiles shortly after the
SBH particles had formed a hard binary; thereafter
the nuclear profile evolved slowly toward even shallower slopes
as the massive binary ejected stars.
As Fig. 5 illustrates, the stellar density around the binary
drops very quickly after the binary reaches a separation $a\approx a_h$.

\begin{figure}[h]
  \def\epsfsize#1#2{0.8#1}
  \centerline{\epsfbox{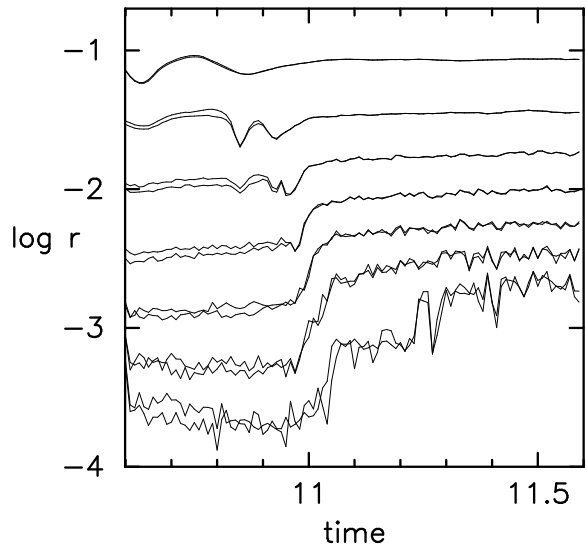}}
  \caption{\it 
Lagrangian radii around each of the two SBH particles 
in an equal-mass merger simulation \cite{mm-01}.
>From bottom to top, the radii enclose 
$10^{-4}$, $10^{-3.5}$, $10^{-3}$, $10^{-2.5}$, 
$10^{-2}$, $10^{-1.5}$ and $10^{-1}$ in units of the mass of one galaxy before the merger.
The binary becomes ``hard'' at $t\approx 11$, and very rapidly
heats the surrounding stellar fluid, lowering the local density.}
  \label{fig:lagrange}
\end{figure}

The hardening rate of the binary in these simulations was found {\it not} to 
be strongly dependent on the number of particles.
This result was subsequently shown \cite{mm-03b} to be due to the small $N$: 
stars were resupplied to the loss cone via collisions
at a higher rate than they were being kicked out by the
binary, ensuring a continuous supply of stars and allowing
the binary to continue to shrink.
While a qualitatively similar evolution may take place in some
galaxies -- for instance, loss cones in non-axisymmetric potentials
can be continuously repopulated by stars on centrophilic
orbits (\S 4.4) -- {\it collisional} loss cone refilling
is very unlikely to achieve anything like a full loss cone
except in very small, dense galaxies.
The long-term evolution of the binary in almost all published $N$-body
simulations are therefore not representative of what one would expect 
in real galaxies.

\begin{figure}[h]
  \def\epsfsize#1#2{0.65#1}
  \centerline{\epsfbox{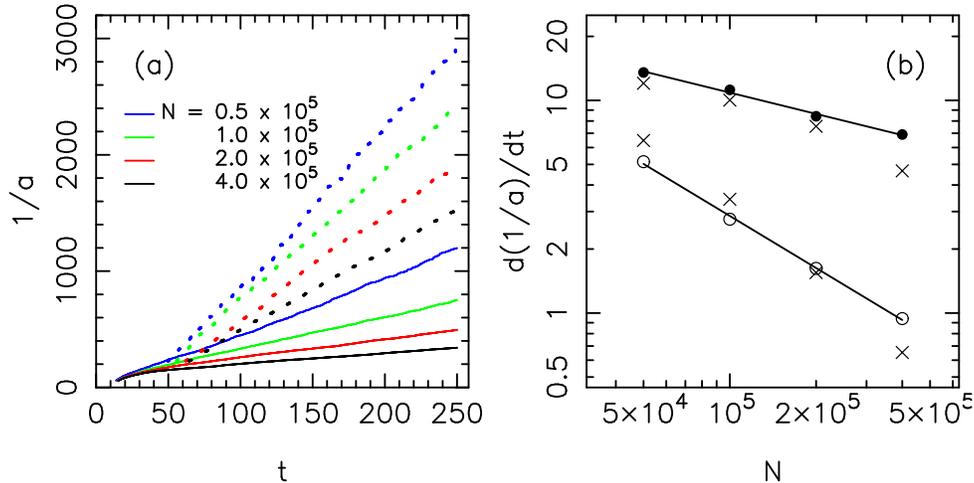}}
  \caption{\it Evolution of the binary semi-major axis (a) and
    hardening rate (b) in a set of high accuracy $N$-body simulations; 
    the initial galaxy model was a low-central-density Plummer sphere
    (\cite{berczik-05}).
    Units are $G=M_{gal}=1$, $E=-1/4$, with $E$ the total energy.
    (a) Dashed lines are simulations with binary mass $M_1=M_2=0.005$
    and solid lines are for $M_1=M_2=0.02$, in units where the
    total galaxy mass is one.
    (b) Filled(open) circles are for $M_1=M_2=0.005(0.02)$.
    Crosses indicate the hardening rate predicted
    by a simple model in which the supply of stars to the binary
    is limited by the rate at which they can be scattered
    into the binary's influence sphere by gravitational encounters.
    The simulations with largest ($M_1,M_2$) exhibit the
    nearly $N^{-1}$ dependence expected in the ``empty loss cone''
    regime that is characteristic of real galaxies.
  \label{fig:berczik}
}
\end{figure}

The relevant dimensionless parameter is
\begin{equation}
q(E) \equiv {(\delta J)^2\over J_{lc}^2},
\end{equation}
where $\delta J$ is the change over one radial period in the
angular momentum of a star on a low-$J$ orbit, and $J_{lc}$
is the angular momentum of an orbit at the edge of the 
binary's loss cone.
A value $q(E)\gg 1$ implies that the loss cone orbits at energy
$E$ are re-populated at a much higher rate than they are 
de-populated by the binary, and the loss cone remains
nearly full.
A value $q\ll 1$ implies that the loss cone is essentially
empty, and repopulation must talke place diffusively,
as stars scatter in from $J\gap J_{lc}$.
In real galaxies, $N$ is large and $\delta J$ is small,
implying $q \ll 1$.
Achieving $q\ll 1$ in $N$-body simulations requires
large particle numbers, and/or a model for the galaxy
that has an unrealistically low central density, so that
the star-star relaxation time is long.
Figure~\ref{fig:berczik} shows a recent set of $N$-body simulations
that does both \cite{berczik-05}.
The massive binary was embedded in a Plummer \cite{plummer-11} galaxy
model, which has a core radius comparable to its half-mass radius;
this model is very different from real galaxies but its very
low degree of central concentration implies a long relaxation
time and low rate of collisional loss-cone refilling.
Large particle numbers were achieved, without sacrificing
accuracy, by running the simulation on a parallel GRAPE cluster.
The $N$-dependence of the binary's hardening rate is clear;
in the simulations with binary mass $M_1=M_2=0.02 M_{gal}$, 
the $N$-dependence
of the hardening rate is $s\equiv (d/dt)(1/a)\approx N^{-0.8}$,
almost as steep as the $N^{-1}$ dependence predicted for an
``empty'' loss cone \cite{mm-03b}.
A similar study \cite{makino-04}, based on King-model galaxies,
found a similar result.

The Plummer-model initial conditions used in the simulations
of Figure~\ref{fig:berczik} were identical to those
adopted in two other $N$-body studies based on a more
approximate $N$-body code \cite{quinlan-97,chatterjee-03}.
Contrary to Figure~\ref{fig:berczik}, Chatterjee, Hernquist
\& Loeb (2003) found that the binary hardening rate
``saturated'' at values of $N\gap 2\times 10^5$, remaining
constant up to $N\approx 4\times 10^5$.
They speculated that this was due to a kind of Brownian-motion-mediated
feedback, in which the binary maintains a constant supply
rate by modulating the local density of stars.
However no supporting evidence for this model was presented;
for instance, it was not demonstrated that the central density
was actually regulated by the binary, or that the amplitude
of the Brownian wandering increased with $N$ in the manner postulated.
Furthermore these authors provided no plots showing the
claimed $N$-dependence of the hardening rate.
Chatterjee et al.'s conclusion, 
that ``a substantial fraction of all massive binaries 
in galaxies can coalesce within a Hubble time,'' is not substantiated
by the more accurate $N$-body simulations shown in Figure~\ref{fig:berczik}.

While Brownian motion probably does affect the decay rate of
binaries in $N$-body simulations \cite{mm-03b}, it is doubtful
that the effect is significant in real galaxies.
The Brownian velocity of {\it single} black holes is found 
in $N$-body integrations to be
\cite{laun-04}
\begin{equation}
{1\over 2}M\langle V^2\rangle \approx {3\over 2}m_\star\tilde\sigma^2
\label{eq:brown}
\end{equation}
where $\tilde\sigma^2$ is the 1D, mean square stellar velocity within
a region $r\lap 0.5 r_h$ around the black hole (and includes the
influence of the black hole on the stellar motions), and $m_\star$
is the stellar mass.
In the case of the Milky Way black hole, equation (\ref{eq:brown})
implies $V_{\rm rms}\approx 0.2$ km s$^{-1}$ (assuming $m_\star=M_\odot$)
and an rms displacement of $\lap 0.1$ pc. 
Brownian motion of a massive {\it binary} is larger than
that of a single black hole, but only by a modest factor
\cite{merritt-01,mm-01,makino-04}.
The rms displacement of a binary from its otherwise central
location would therefore be very small in a real galaxy, probably even
less than the separation between the two components of the binary.

\begin{figure}[h]
  \def\epsfsize#1#2{0.65#1}
  \centerline{\epsfbox{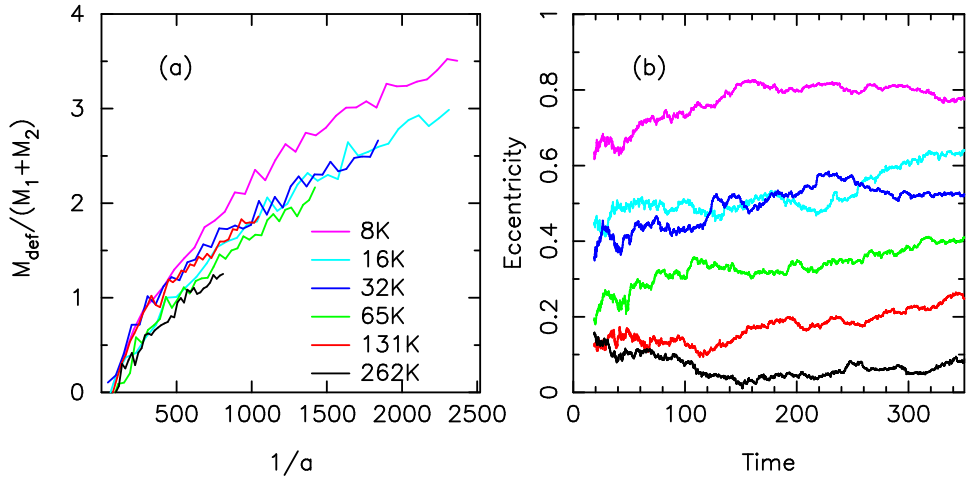}}
  \caption{\it Results from a set of $N$-body integrations
    of a massive binary in a galaxy with a $\rho\sim r^{-0.5}$
    density cusp \cite{szell-05}.
    Each curve is the average of a set of integrations
    starting from different random realizations of the same
    initial conditions.
    (a) Evolution of the ``mass deficit'' (Eq.~\ref{eq:mdef}),
    i.e. the mass in stars ejected by the binary.
    For a given value of binary separation $a$,
    the mass deficit is nearly independent of particle
    number $N$, implying that one can draw conclusions
    from observed mass deficits about the binary
    that produced them.
    (b) Evolution of binary eccentricity.
    The eccentricity evolution is strongly $N$-dependent
    and tends to decrease with increasing $N$, suggesting
    that the eccentricity evolution in real binaries would
    be modest.
  \label{fig:szell}
}
\end{figure}

The goal of $N$-body studies is to simulate binary evolution
in galaxies with realistic density profiles, and with large enough
particle numbers that re-population of the binary's loss cone
takes place diffusively, as in real galaxies.
Two avenues are open for making further progress in this area.
1. One could combine a Monte-Carlo treatment of stellar encounters
\cite{freitag-01,freitag-02} with a lookup table, derived from
scattering experiments, of energy and angular momentum changes 
experienced during close passages of stars to the binary
\cite{quinlan-97,merritt-01}.
Such a hybrid algorithm would allow one to adjust the degree
of collisionality at will and record the effects on both
the binary's evolution, and the influence of the binary
on the stellar distribution.
This approach would be difficult to generalize to non-spherical
geometries however.
2. A straightforward $N$-body approach is also feasible,
but particle numbers in excess of $\sim 10^7$ are required
\cite{mm-03b}.
Such large particle numbers are just now becoming feasible for
direct-summation $N$-body codes,
by combining the GRAPE accelerator boards \cite{namura-03}
with a parallel architecture \cite{dorband-03}.
Indeed just such an approach was used for the integrations
of Figure~\ref{fig:berczik}, although the galaxy models
in that study were rather unphysical.

Figure~\ref{fig:szell} shows a promising early step in this direction.
The initial galaxy models had $\rho\sim r^{-0.5}$ density cusps;
integrations were carried out on a GRAPE-6 computer, limiting
the total particle number to $\sim 256K$, but  the motion
of the black hole binary (of mass $M_1=M_2=0.005 M_{gal}$)
and nearby stars was carried out
using the Mikkola-Aarseth chain regularization algorithm
\cite{mikkola-90,mikkola-93,aarseth-03b}.
Because of the models' higher central density and limited
particle numbers, the binary's loss cone was only partially
empty, $q\gap 1$, and the $N$-dependence of the hardening
rate was shallower than expected in real galaxies,
$(d/dt)(1/a)\sim N^{-0.4}$.
Figure~\ref{fig:szell} (a) shows that the ``damage'' inflicted
by the binary on the nucleus is {\it not} strongly dependent
on $N$, as expected (cf. Eq.~\ref{eq:def_J}).
This is an encouraging result since it implies that one
can hope to learn something definite about pre-existing
binaries by comparing $N$-body simulations with observations 
of the centers of current-day galaxies (\S 7).
On the other hand, Figure~\ref{fig:szell} (b) suggests
that the evolution of the binary's eccentricity {\it is}
strongly $N$-dependent.
This may explain the rather disparate results on eccentricity
evolution in past $N$-body studies
\cite{mm-01,hemsendorf-02,aarseth-03}.

Much progress on this problem is expected in the next few years.

\section{Evidence for Cusp Destruction}
\label{sec:cusp}

A potentially powerful constraint on models of binary SBH
evolution is the observed central density structure of galaxies.
Figure~\ref{fig:mej} shows that a massive binary
must eject of order its own mass in reaching a separation
at which $t_{gr}\lap 10^{10}$ yr if $m_2\approx m_1$, 
or several times $m_2$ if $m_2\ll m_1$.
These numbers should be interpreted with caution since:
(1) binaries might not decay this far -- they may stall -- or
the final stages of decay might be driven by gas dynamics
rather than energy exchange with stars; (2) the definition of
``ejection'' used in Figure~\ref{fig:mej} is escape of a star
from an isolated binary, and does not take into account the
confining effect of the nuclear potential; 
(3) the effect of repeated mergers on nuclear density profiles, 
particularly mergers involving very unequal-mass binaries, 
is poorly understood.
Neverthless, even the initial formation of a hard binary
displaces a mass of order $m_2$ (Fig.~\ref{fig:lagrange}).

The luminosity profile data can probably be used to rule
out one model of binary evolution.
In a ``collisionless'' galaxy (Table 1), the binary's
loss cone never refills, and decay of the binary would stall.
The binary carves out a ``hole'' in both phase
space and configuration space; the radius of the latter would
be $\sim 3a_h$ \cite{zier-02}.
While central minima may have been seen in the luminosity
profiles of a few galaxies \cite{lauer-02}, these are
likely due to dust obscuration, and the great majority
of galaxies show a clearly rising stellar density
into radii $\lap r_{\rm infl}$.
The non-existence of true ``cores'' suggests either that some
degree of loss-cone refilling occurs, or that the final
decay of the binary takes place via a more efficient process
than ejection of stars.

Nevertheless there is a well-defined trend for the central
densities of bright galaxies to decrease
with increasing luminosity 
\cite{ferrarese-94,mf-95,faber-97,gebhardt-96,graham-03}.
Nuclear densities in elliptical galaxies and spiral bulges 
with $M_V\lap -20$ follow $\rho\sim r^{-\gamma}$,
$\gamma\lap 1$, while in fainter spheroids, $1\lap\gamma\lap 2.5$.
A natural interpretation is that the brightest galaxies -- which
presumably formed via one or more mergers -- have experienced
more cusp destruction than fainter galaxies.
(An alternative possibility, discussed below, is that the nuclei in
faint galaxies re-formed after being destroyed.)

In practice, this hypothesis is difficult to test since it requires
knowledge of the pre-merger density profiles.
A reasonable guess is that all galaxies originally had steep
power-law density cusps, since these are generic in the 
faintest galaxies known to harbor SBHs.
For instance, both M32 and the bulge of the Milky Way
have $\rho \sim r^{-1.5}$ at $r\lap r_{\rm infl}$
and $\rho\sim r^{-2}$ just outside
\cite{lauer-98,genzel-03}.

\begin{figure}[h]
  \def\epsfsize#1#2{0.7#1}
  \centerline{\epsfbox{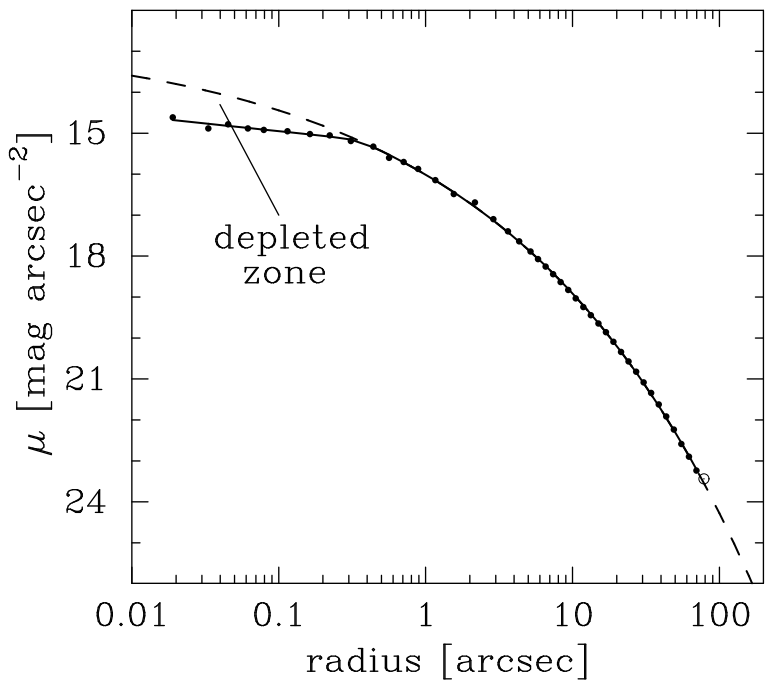}}
  \caption{\it Observed surface brightness profile of NGC 3348.
  The dashed line is the best-fitting Sersic model to the large-radius
  data.  Solid line is the fit of an alternative model,
  the ``core-Sersic'' model, which fits both the inner and outer
  data well. The mass deficit is illustrated by the area
  designated ``depleted zone'' and the corresponding mass is
  roughly $3\times 10^8\msun$ \cite{graham-04}.}
  \label{fig:sersic}
\end{figure}

The ``mass deficit'' \cite{milos-02} is defined as the difference
in integrated mass between the observed density profile
and the primordial (pre-merger) profile.
For instance, if the primoridal profile is a power law
of index $\gamma_0$ inward of some radius $r_b$, then
\begin{equation}
M_{\rm def} \equiv 4\pi\int_0^{r_b}\left[\rho(r_b)\left({r\over r_b}\right)^{-\gamma_0} - \rho(r)\right]r^2dr.
\label{eq:mdef}
\end{equation}
Mass deficits in samples of bright elliptical galaxies
were computed in three recent studies 
\cite{milos-02,ravin-02,graham-04}.
In the first two studies, the authors assumed power-laws
of various slopes for the pre-merger profiles, and found
$\langle M_{\rm def}/M_\bullet\rangle\approx 1$ 
for $\gamma_0=1.5$ with $M_\bullet$ the current mass of the SBH.
The latter study made use of the fact that the light profiles
of bright galaxies show an abrupt downward deviation relative to
a Sersic \cite{sersic-68} profile fit to the outer regions
(Fig.~\ref{fig:sersic}).
Mass deficits inferred in this study were slightly larger, 
$M_{\rm def}/M_\bullet\approx 2$.\footnote{The author of this 
study presents his mass deficits as significantly {\it smaller}
than those found in the earlier studies.
However he bases his comparison on values of $M_{\rm def}$
computed exclusively using $\gamma_0=2$.}
These numbers are within the range predicted by the binary
SBH model, particularly given the uncertainties associated with
the effects of multiple mergers.

In small dense galaxies, a destroyed cusp would be expected to
re-form via the Bahcall-Wolf \cite{bw-76,preto-04} process, on 
a timescale of order the star-star relaxation time measured
at $r_{\rm infl}$.
This time is of order $10^9$ yr in the Milky Way bulge
and the nucleus of M32.
This may be the explanation for the steep power-law profiles
observed at the centers of these galaxies.
Alternatively, the steep cusps may be due to star formation
that occurred after the most recent merger \cite{hk-00}.

\begin{figure}[h]
  \def\epsfsize#1#2{0.6#1}
  \centerline{\epsfbox{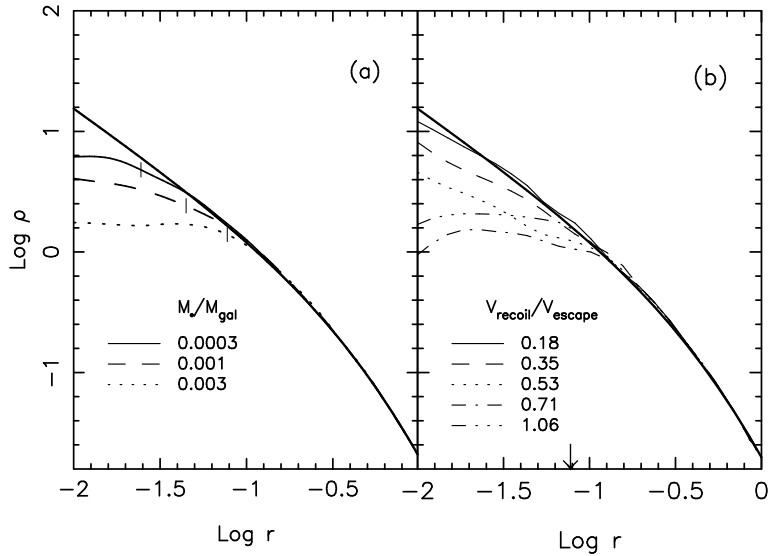}}
  \caption{\it Effect on the nuclear density profile of SBH ejection.
The initial galaxy model (black line) has a $\rho\sim r^{-1}$
density cusp.
(a) Impulsive removal of the SBH.
Tick marks show the radius of the black hole's sphere of influence
$r_{\rm infl}$ before ejection.
A core forms with radius $\sim 2r_{\rm infl}$.
(b) Ejection at velocities less than escape velocity.
The black hole has mass $0.3\%$ that of the galaxy;
the galaxy is initially spherical and the black hole's
orbit remains nearly radial as it decays via dynamical friction.
The arrow in this panel marks $r_{\rm infl}$ in the initial
galaxy. \cite{mmfhh-04}.}
  \label{fig:eject}
\end{figure}

More rigorous tests of the binary SBH model will require a
better understanding of the expected effect of massive binaries on
stellar density profiles.
As discussed above, while the best current $N$-body simulations
suggest $\rho\sim r^{-1}$ following binary formation \cite{mm-01},
the simulations are dominated by noise over the long term.

A number of other processes could compete with binary SBHs
in the destruction of nuclear density cusps.
A population of three or more SBHs in a galactic nucleus
would undergo a complicated set of close encounters resulting
ultimately in coalescence and/or ejection of some or all of the SBHs
(\S 5).
In the process, the stellar background would be heated and
a mass of order five times the
combined mass in SBHs removed \cite{mm-02}.
This model reproduces the observed time dependence
of core radii in globular clusters very well \cite{merritt-04b}
but its relevance to galactic nuclei is less clear;
the model requires binary coalescence times long enough
that an uncoalesced binary is present when a third SBH falls in
\cite{volonteri-03}.
If a binary SBH does eventually coalesce, 
the gravitational radiation carries a 
linear momentum leading to a recoil of the 
coalesced hole \cite{bekenstein-73,fitchett-83}.
Recoil velocities are estimated to be as large as
$\sim 400$ km s$^{-1}$ \cite{favata-04,mmfhh-04},
although with considerable uncertainty.
A SBH ejected from a galactic nucleus with 
a velocity of $\sim 10^2$ km s$^{-1}$ 
would quickly fall back to the center,
but its displacement and infall would heat the 
stellar fluid and lower its density.
Figure~\ref{fig:eject} 
shows the effects of ejection on nuclear density
profiles.
Mass deficits produced by this mechanism can be comparable
in amplitude to those predicted by the binary SBH model.

A major focus of future work should be to calculate
the evolution of $\rho(r)$ as predicted by the various
scenarios for binary decay discussed in this article.

%\newpage

%===================================================================================

\section{Interaction of Binary Black Holes with Gas}
\label{sec:gas}

Interstellar gas might play an important role in the dynamical 
evolution of a binary SBH. 
Interactions with gas complement interactions with the stellar 
environment (\S\ref{sec:stars}) and with other SBHs (\S\ref{sec:mult}). 
Any gas situated close to a binary is disturbed by the SBHs and exerts 
gravitational force on them, thereby affecting their orbit.  
Furthermore, if SBH coalescence is accompanied by the presence of gas, 
an observable electromagnetic afterglow might follow coalescence. 

The collisional, dissipative nature of interstellar gas 
gives rise to a behavior fundamentally different from that of the point-mass dynamics of stellar systems.  
It is natural to distinguish between two classes of flows in dynamical systems containing gas. 
In {\it hot flows} the gas temperature is comparable to the 
virial temperature of the system, 
while in {\it cold flows} the gas temperature is significantly 
below the virial temperature.  
The virial temperature can be defined as $T_{\rm vir}=GM_{12}\mu m_p/2ak$, where $\mu$ is the mean particle mass in units of the proton mass $m_p$, and $k$ is the Boltzmann constant.  
The prototype of a hot flow is the spherical, ``Bondi'' accretion onto a single black hole, in which the accreting gas is supported by pressure against free infall toward the accretor.  
The prototype of a cold flow is a thin disk, in which the gas is rotationally supported against infall.  Even in hot flows rotational support is realized close to the accretor when the gas has nonzero net angular momentum (e.g.~\cite{krumholz-04}).  

The angular momentum barrier is central to SBH formation theories.  
Any model for how material is channeled into an accreting black hole must describe the mechanism by which angular momentum is removed from the material. Whatever this mechanism may be, it is expected that it operates universally during the epoch in which SBHs grew to their present masses by rapidly accreting material onto pre-existing black hole ``seeds.''  This is also the period when galaxy merging peaks \cite{hk-00,haehnelt-02,wyithe-02,volonteri-03b}.  While still elusive to astronomical probes due to severe obscuration \cite{sanders-88}, the nuclei of merging galaxies, which are also the sites for the formation of binary SBHs \cite{bbr-80}, are expected to contain the largest concentration of dense gas anywhere in the universe.  
The inevitable abundance of gas motivates an inquiry into the role of gas dynamics as an alternative to stellar dynamics in the process of SBH coalescence.  
Some of the mechanisms that remove angular momentum of interstellar gas and thus channel it into the neighborhood of SBHs include the torquing of gas flow by the rapidly-fluctuating potential of merging galaxies \cite{mihos-96} and by nested stellar bars \cite{sfb-89}, angular momentum transport by hydrodynamical turbulence that might be driven by the onset of self-gravity \cite{shlosman-89,gammie-01,goodman-01} or by supernovae embedded within a large-scale toroidal circumnuclear flow \cite{wada-04}, 
angular momentum extraction by magnetohydrodynamical turbulence \cite{balbus-98} or by magnetic braking \cite{blandford-82}, and more speculatively, by Rossby vortex instabilities \cite{li-00}.

Astronomical observations offer abundant evidence for both hot and cold gas flows in the immediate vicinity of SBH candidates.  The origin and the dynamical impact of the two classes of gas flow are distinct and are discussed here separately.

\subsection{Interaction with Hot Gas}

Hot gas permeates the interstellar space in galaxies and the intergalactic space in groups of galaxies and galaxy clusters.  
Virial temperatures range between $10^6\textrm{ K}-10^8\textrm{ K}$ and the 
hot gas is almost completely ionized. 
Primordial and secondary sources contribute to the pool of hot has.  During the early stages of galaxy formation, intergalactic space contains partially ionized gas inherited from the pregalactic, early universe.  Hydrogen recombines at redshifts $z\sim 1000$ and is reionized at redshifts $z\sim 10$ by the radiation emitted by the earliest structures.  
The partially ionized gas cools within the confining gravitational potential of dark matter halos and filaments. 
Cold gas accelerates toward the halos' centers of gravity and is shock-heated to about the virial temperature.  
Some of the coldest inflowing gas escapes heating by accreting along narrow channels that reach deep inside the primary halo.  Cooling times in the halo centers where the gas is the densest are short compared to the dynamical time and thus most of the primordial gas is consumed in starbursts on a dynamical time scale.  

Tenuous gas that remains after the cooling time has exceeded the dynamical time in the nascent galaxy might still be plentiful enough to feed a massive black hole growing at an Eddington-limited rate.  
The residual number density at the radius of influence of the SBH is
\begin{eqnarray}
n &\approx& \frac{\sigma^3 kT}{G M \Lambda} \nonumber\\
&\approx& 20\textrm{ cm}^{-3} \mu
\left(\frac{M}{10^8M_\odot}\right)^{0.11}
\left(
\frac{\Lambda}{2\times10^{-23}
\textrm{erg cm}^{3}\textrm{ s}^{-1}} 
\right)^{-1} ,
\end{eqnarray}
where $T$ is the virial temperature of the galaxy, 
$\mu$ is the average atomic mass in units of the proton mass, 
$\Lambda$ is the cooling function \cite{dalgarno-72}, and we have employed the $M$ -- $\sigma$ relation (Eq. \ref{eq:ms}) to relate the virial temperature to the black hole mass.  The thermal stability limit could in principle be exceeded if the gas kept at the Compton temperature by a continuum flux from an unobscured AGN \cite{fabian-90}. 

This so-called ``cooling flow model of quasar fueling'' 
\cite{ciotti-97,nulsen-00} is however plagued by many problems 
(\cite{krolik-99} and references therein).  Most of the gas left over from star formation might be blown out by the mechanical feedback associated with the radiative and mechanical output of the accreting massive black hole \cite{silk-98,king-03,murray-04}.  A small amount of angular momentum in the gas results in circularization and settling into an accretion disk. This disk may be susceptible to fragmentation, thereby converting most of the gas mass into stars and effectively cutting off the supply of gas to the SBH \cite{tan-04}.

The geometry of the flow of a hot, magnetized gas near a binary black hole is unknown.  
Assuming spherical, non-rotating accretion, 
the time scale on which the hot gas is captured by the SBH is
\begin{eqnarray}
t_{\rm capt}&\equiv& \frac{M}{\dot M} \nonumber\\
&\approx& f_{\rm b}\frac{\sigma^3}{ G^2 M \mu m_p n}\nonumber\\
&\approx& 10^8\textrm{ yr}\ f_{\rm b }\mu^{-2}\left(\frac{M}{10^8M_\odot}\right)^{-0.44}
\left(\frac{\Lambda}{2\times10^{-23}
\textrm{erg cm}^{3}\textrm{ s}^{-1}}\right) ,
\end{eqnarray} 
where $f_{\rm b}\sim 1-10$ is a numerical factor that depends 
on the equation of state of the gas. 

If a binary black hole is present, gravitational torques 
from the gas induce decay of the binary's semi-major axies
on approximately the same time scale.  
This crude estimate is based on an analogy with binary-star interactions: 
the binary must eject of order its own mass in stars to decay an $e$-folding 
in separation.  
Hot gas torquing the binary might be ejected in an outflow and thus the actual rate at which gas is accreting onto individual binary components might be severely suppressed compared to the accretion expected in an isolated black hole.

Galactic nuclei also contain hot gas produced by secondary sources. 
For example, observations with the {\it Chandra} X-ray Observatory 
have revealed tenuous ($n\approx 10-100\textrm{ cm}^{-3}$), 
hot ($T\approx1\textrm{ keV}$) plasma within a parsec of the 
$\sim4\times10^6M_\odot$ Milky Way SBH \cite{baganoff-03}. 
This plasma is being generated by the numerous massive, evolved stars in the galactic region \cite{genzel-03} through stellar wind and supernova activity.  Since its temperature is higher than the virial, most ($>99\%$) of the plasma escapes the neighborhood of the SBH \cite{quataert-99}.  
While the hot gas densities in active galaxies might be transiently 
larger than that at the Galactic center, 
the tendency of the hot plasma to escape the neighborhood of the 
SBH reduces the likelihood that large quantities of virialized gas 
would remain enmeshed with the binary's orbit long enough to affect its dynamical evolution.

Recently, Escala et al.~\cite{escala-04,escala-05} carried out smoothed 
particle hydrodynamical (SPH) simulations of binary point masses 
interacting with a massive, spherical cloud of hot gas initially 
centered on the binary. Gravitational drag from the gas induces 
decay in the binary's orbit.  The relevance of spherical, hot 
initial conditions is contingent on the astrophysical plausibility 
that a compressed accumulation of hot gas comparable in mass to the 
SBH can be sustained.

\subsection{Interaction with Cold Gas}

The specific angular momentum of a cold flow
might easily exceed that of the binary. 
The gas then tends to settle into rotationally
supported, geometrically thin rings and disks (recall that ``cold'' gas is colder than the virial temperature but can be hot enough to be ionized).  

Observations offer abundant evidence for the presence 
of dense gas in galactic nuclei.
Thin, Keplerian 
molecular disks on
scales $0.1\textrm{ pc}-0.5\textrm{ pc}$ 
have been seen in the water maser
emission in the nuclei of Seyfert galaxies
\cite{miyoshi-95,gallimore-01,greenhill-03}.  The Galactic nucleus contains a $4\times10^6 M_\odot$ black hole surrounded by 
a $\sim(10^4-10^5)M_\odot$ molecular gas torus at distances 
$>1\textrm{ pc}$ from the SBH \cite{jackson-93}.  
Compact stellar disks on
scales $\geq20\textrm{ pc}$, which are fossil evidence of past gas
circularization, are evident in the nuclei are of many galaxies
\cite{pizzella-02}.  Massive accretion disks must be 
present in quasars and the Narrow-Line Seyfert I nuclei to account
for what appears to be rapid accretion onto the central SBHs in these systems. 
However, the structure of these
disks at radii comparable to the size of a hard SBH binary is unknown.  
The principal challenges to constructing extended disk models are the instabilities related to incomplete ionization and the susceptibility to gravitational fragmentation \cite{kolykhalov-80,shlosman-89,milosavljevic-04,goodman-04}. 

If a disk surrounding a binary SBH is initially inclined with respect 
to the binary's orbital plane, 
the quadrupole component of the binary's gravitational potential causes differential precession in the disk at the rate \cite{larwood-96} 
\begin{equation}
\Omega_{\rm prec}(r)=\frac{3}{4}\frac{q}{(1+q)^2} \frac{(GM_{12})^{1/2}a^2}{r^{7/2}} ,
\end{equation}
which results in a warping of the disk.  
As in the Bardeen-Petterson mechanism 
\cite{bardeen-75,pringle-92,scheuer-96}, 
the warp either dissipates, or smears around the binary, 
resulting ultimately in a nearly axisymmetric disk in the binary's 
orbital plane.  

Interest in co-planar, circumbinary disks stems from their ability to 
extract a binary's angular momentum via a form of tidal coupling.  
Two interrelated questions might be posed: What is the response of a circumbinary disk to the binary's tidal forcing? and: 
How does such a disk affect the evolution of the binary's orbit?

Existing attempts to answer these questions have employed ad hoc models for the form of the binary-disk torque coupling \cite{pringle-91,ivanov-99}, or have been restricted to binaries with components of very unequal mass \cite{armitage-02}, where an array of neighboring Lindblad resonances facilitate binary-disk coupling \cite{goldreich-79}, much like the coupling between a massive planet and its natal gas disk \cite{goldreich-80}.  Early numerical simulations of circumbinary disks with nearly equal masses \cite{artymowicz-94}, however, suggested that the disks are truncated exterior to the resonances, which was interpreted as a consequence of a collisionless nonlinear parametric instability \cite{rudak-81,erwin-99}.  Fluid dynamical theory of circumbinary disk truncation is still lacking.   

In a circular binary the outer Lindblad resonances (OLR) are located at radii $r_m=(1+1/m)^{2/3}$, where $m=1,2,...$ is the order in the decomposition of the binary's gravitational potential into multipoles:
\begin{equation}
\varphi(r,\theta)=\sum_{m=0}^\infty \varphi_m(r) \cos[m(\theta-\Omega_{\rm bin}t)] .
\end{equation} 
The outermost OLR is located at $r\approx 1.6 a$.  The resonances are radii in the disk where the natural, epicyclic frequency of radial oscillations in the disk is an integer multiple of the rate at which a packet of disk gas receives tidal ``kicks'' by the binary.  The forcing near a resonance, as well as at a radius where surface density in the disk exhibits a large gradient, 
excites nonaxisymmetric propagating disturbances, or ``density waves,'' in the disk.

The gravitational potential of eccentric binaries contains low-frequency 
components that are absent in circular binaries.  
These low-frequency components activate resonances located at larger radii than in the circular case, and might lead to mutual excitation and reinforcement of the binary and the disk eccentricities \cite{papaloizou-01,goldreich-03}.  Many extrasolar planets, which are thought to form in circumstellar disks, are notably eccentric\footnote{http://exoplanets.org}, suggesting that dynamical coupling between a binary point mass (a star and a planet, or a pair of black holes) and a gas disk is conducive to eccentricity growth. The observed circumbinary disks in young stellar binaries such as GG Tau, which are typically eccentric, 
are truncated at radii a few times the semimajor axis \cite{mccabe-02}, 
which lends support to this hypothesis.  
Eccentricity in SBH binaries accelerates coalescence due to gravitational
wave emission (Eq.~\ref{eq:peters0}) and might be detectable in gravitational wave trains.  

Density waves transport angular momentum outward through the circumbinary disk.  Angular momentum flux carried by the waves is extracted from the binary's angular momentum. 
The binary experiences a negative torque equal and opposite to the total 
angular momentum flux transferred to the disk.  
The location of the inner edge of the disk reflects a balance 
between the angular momentum flux deposited into the disk, 
and the angular momentum flux transported through the disk by another, 
possibly viscous mechanism.  
Wave momentum is deposited into the disk material via a form of 
dissipative damping. 
The location in the disk where the waves are damped can be 
separated by many wavelengths from the location where they are excited. 
The damping could take place in the nonlinear steepening and the 
breaking of wave crests \cite{savonije-94,rafikov-02}.  
In marginally optically thick disks, radiation damping might 
also play a role \cite{cassen-96}.  
Yet another form of damping could be due to the dissipation of 
wave shear if the disk is strongly viscous \cite{takeuchi-96}.  
The amplitude of the density waves is a steeply decreasing function 
of the radius of excitation. 
The amplitude is diminished if the waves are nonlinear at excitation 
and damp in situ, but then one expects the inner edge to recede 
where in situ damping shuts off.  

The intricate and insufficiently understood nature of binary-disk interactions calls for grid-based hydrodynamical simulations with a shock-capturing capability.  The necessity that the radial wavelength, which is smaller than the vertical scale height of the disk, be resolved by multiple cells, 
places severe demands on the computational resources, 
especially if a three dimensional representation of the disk is required. 
It should also be noted that the radiative and thermal structure of accretion 
disks around {\it single} SBHs are not adequately understood on any 
radial scale.

As a binary's semimajor axis decreases due to stellar, gas dynamical, 
or gravitational radiation processes, a circumbinary disk's inner edge 
spreads inward viscously while maintaining constant edge-to-semimajor 
axis ratio, e.g., $r_{\rm edge}/a\sim 2$.  
In the final stages of the gravitational radiation-driven inspiral, however, the time scale on which the semimajor axis decays becomes shorter than the viscous time scale, and the disk can no longer keep up with the binary, resulting in binary-disk detachment.  On the relevant length scales the disk might be dominated by radiation pressure and the electron scattering opacity; the structure and the stability of such disks is an active research area \cite{turner-04}.

%\newpage

%===================================================================================

\section{Spin Evolution during Mergers}
\label{sec:spin}

Coalescence of a binary black hole results in a spinning 
remnant.\footnote{In this section, we set $G=c=1$.}
Angular momentum conservation implies
\begin{equation}
{\bf S}_1 + {\bf S}_2 + {\bf L}_{\rm orb} = {\bf S} + {\bf J}_{\rm rad}
\end{equation}
where ${\bf S}_1$ and ${\bf S}_2$ are the spin angular momenta of the two
SBHs just before the final plunge, ${\bf L}_{\rm orb}$ is the orbital
angular momentum of the binary before the plunge, 
${\bf S}$ is the spin of the resulting black hole, and
${\bf J}_{\rm rad}$ is the angular momentum carried away
by gravitational waves during and after the coalescence~\cite{flanagan-98}.
The simplest case to treat is extreme mass ratio mergers, 
$q\equiv m_2/m_1\ll 1$, for which 
the binary can be described as a test particle of mass $m_2\equiv m$
orbiting a black hole of mass $m_1\equiv M\gg m$, 
and both ${\bf S}_2$ and ${\bf J}_{\rm rad}$ 
can be ignored.
The change in the larger hole's spin is computed
by adding the smaller hole's energy and orbital angular momentum at 
the last stable orbit (LSO).
The latter varies from $L_{\rm LSO}/m=\sqrt{12}M$ for circular
equatorial orbits around a non-spinning
hole to $L_{\rm LSO}=M(9M)$ for prograde(retrograde) orbits around a 
maximally-spinning hole, ${\bf S}_1=M^2$.
The much larger value of $L_{\rm LSO}$ in the case of retrograde
capture implies that a rapidly-rotating hole will typically spin
down if capture occurs from random directions 
\cite{doroshkevich-66,godfrey-70,young-76,young-77,hughes-03}.
The change in spin assuming $q\ll 1$ is
\begin{equation}
\delta\hat a = q\left(- 2\hat a + {\hat L}_{\rm LSO,z}\right) + O(q^2)
\label{eq:deltaa}
\end{equation}
where $\hat a\equiv |{\bf S}_1|/M^2$, $L_z$ is the orbital angular
momentum parallel to ${\bf S}$, and $\hat L\equiv L/mM$.
The first term in equation~(\ref{eq:deltaa}) describes conservation
of spin angular momentum of the larger hole as its mass grows,
$\hat a \propto M^{-2}$, while the second term describes the
increase in spin due to torquing by the smaller body.

\begin{figure}[h]
  \def\epsfsize#1#2{0.5#1}
  \centerline{\epsfbox{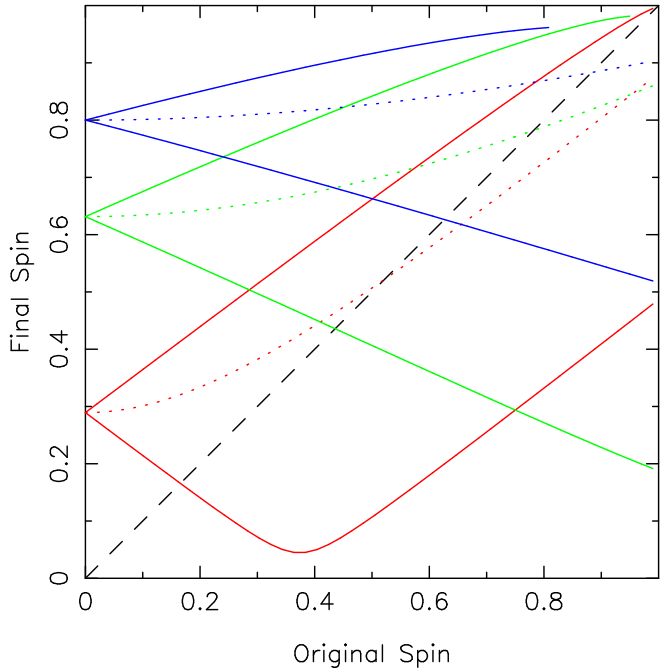}}
  \caption{\it Final spin $\hat a$ of a remnant black hole
    in terms of its
    original spin, for mass ratios $q=0.1$ (red), $0.3$ (green) 
    and $0.5$ (blue).  The change in spin was computed using the
    test-particle approximation for $L_{\rm LSO}$ 
    \cite{bardeen-72,hughes-03,young-77}.
    Upper (lower) curves correspond to prograde(retrograde)
  capture from the equatorial plane; 
  dashed curves are for capture over the pole.
  Capture of a low-mass secondary is likely to spin down the larger
  hole unless the latter is slowly rotating initially.
  Capture of a massive secondary results
  in spinup unless infall is nearly retrograde or the original
  spin is large.}
  \label{fig:deltaa}
\end{figure}

The change in spin after a single coalescence is illustrated
in Fig.~\ref{fig:deltaa} as a function of $q$ and initial spin; 
the upper (lower)
 curves represent prograde (retrograde) captures from equatorial 
orbits,
and the dashed lines are for capture over the pole.
The bias toward spin-down is evident; retrograde capture from
the equatorial plane produces a nearly
(but never completely) non-spinning remnant when 
$q\approx 2.5 \hat a$, $q\le 0.23$, and rapid final rotation
($\hat a\gap 0.9$) requires both a large initial spin and a
favorable inclination. 
On the other hand, if the larger hole is slowly rotating initially,
$\hat a\lap 0.5$, mass ratios $q\gap 0.3$ 
{\it always} result in spin-up.
The oft-repeated statement that ``mergers spin down black holes''
reflects a preconception that SBHs
are likely to be formed in a state of near-maximal rotation
(e.g.,~\cite{bardeen-70,gammie-04}).\footnote{Note the error in Figure 1
of Hughes \& Blandford (2003), which shows the change in spin
for mergers with mass ratio $q=0.5$: the darkest contour on that plot
should be labelled $\hat a = 0.5$, not $\hat a = 0$.}

\begin{figure}[h]
  \def\epsfsize#1#2{0.5#1}
  \centerline{\epsfbox{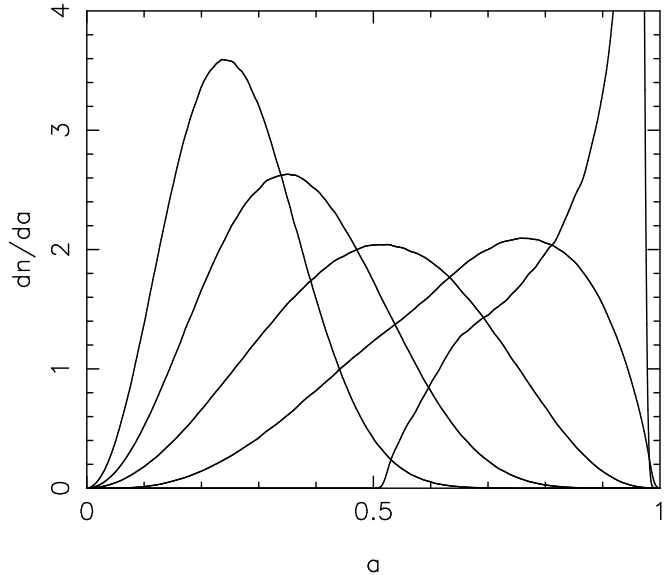}}
  \caption{\it Steady-state spin distributions produced by successive
capture from random directions at fixed mass ratio $q$, for $q=(1/32,1/16,1/8,1/4,1/2)$.
Curves were generated using Monte-Carlo experiments based on the
test-particle approximation, $q\ll 1$; hence
the curve for $q=1/2$ should be viewed as illustrative only.
  }
  \label{fig:nofa}
\end{figure}

Successive mergers from random directions with fixed
$q$ (i.e. secondary mass grows proportionately to primary
mass) lead to a steady-state spin distribution $N(\hat a)$
that is uniquely determined by $q$.
For small $q$, this distribution can be derived from
the Fokker-Planck equation
\cite{hughes-03}:
\begin{equation}
N({\hat a}) d {\hat a} \approx N_0 {\hat a}^2 e^{-3{\hat a}^2/2{\hat a}^2_{\rm rms}} d\hat a,\ \ \ \ \hat a_{\rm rms} \approx 1.58\sqrt{q}.
\label{eq:nofa}
\end{equation}
Figure~\ref{fig:nofa} shows $N(\hat a)$ for various values
of $q$, computed via Monte-Carlo experiments (not
from the Fokker-Planck equation) using the test-mass
approximation for $L_{\rm LSO}$.
The Gaussian form of equation (\ref{eq:nofa}) is seen to be accurate only 
for $q\lap 0.1$.
For $q\gap 1/8$, the distribution is skewed toward large spins.

Accurate calculation of spin-up during a merger of comparably massive
black holes requires a fully general-relativistic numerical treatment.
Adopting various approximations for the radius of the innermost stable
circular orbit (ISCO) for comparably massive binaries
\cite{cook-94,baumgarte-00,grand-02,damour-00},
and assuming that mass and angular momenta are conserved during
coalescence,
gives a remnant spin in equal-mass mergers of
$\hat a\approx 0.8-0.9$.
Baker et al. \cite{baker-02,baker-04} present full numerical calculations
of equal-mass mergers with and without initial spins.
In the absence of initial spins, $3\%$ of the system's mass-energy and
$12\%$ of its angular momentum are lost to gravitational radiation, 
and the final spin is $\hat a\approx 0.72$.
Coalescence of initially spinning holes from circular orbits in the
equatorial plane yields $\hat a \approx 0.72 + 0.32\hat s$ with
$\hat s$ the initial spin parameter of the two holes (assumed equal);
Baker et al. considered initial spins in the range $-0.3\le\hat s\le 0.2$,
where negative/positive values indicate spins aligned/counteraligned
with the orbital angular momentum.
Extrapolating this result toward $\hat s = 1$ suggests that prograde mergers 
of black holes with initial spins $\hat s\ge 0.85$ will result in a 
maximally-spinning remnant.

Confronting these predictions with observation is problematic for
a number of reasons: merger histories of observed SBHs 
are not known,
SBH spins are difficult to determine observationally, 
and other mechanisms, such as gas accretion, can act efficiently
to spin up SBHs \cite{bardeen-70}.
However there is circumstantial evidence that mergers played a 
dominant role in determining the spins of at least some
SBH.
If SBH spins were the product of gas accretion, 
the jets in active galaxies should point
nearly perpendicularly to the disks of their host galaxies
\cite{rees-78}.
In fact, there is almost no correlation between jet direction
and galaxy major axis in Seyfert galaxies \cite{ulvestad-84,
kinney-00,schmitt-02}.
Among the possible explanations \cite{kinney-00} for the misalignment,
perhaps the most natural is that SBH spins were determined
by the same merger events that formed the bulge, long before
the formation of the gaseous disk, and that subsequent spin-up
by gas accretion from the disk plane has been minimal \cite{merritt-02}.
%This argument leaves open the possibility that spin-up during mergers
%took place in part through accretion of gas that was driven to the
%center of the merger remnant.

Coalescence of two black holes during a merger should result in 
a ``spin-flip,'' a reorientation of the spin axis of the
more massive black hole.
In the test-particle limit, the reorientation angle is
\begin{equation}
\delta\theta = {q\over\hat a} {\hat L}_{\rm LSO}\sqrt{1-\mu^2} + O(q^2) 
\end{equation}
with $\mu$ the cosine of the angle between the orbital angular
momentum vector and the spin axis of the larger hole.
When $q\gap 0.2$, the spin orientation is overwhelmed by the plunging
body in a retrograde
merger, even if the initial spin of the larger hole was close to maximal.
Hence, even ``minor mergers'' (defined, following galactic dynamicists, 
as mergers with $q\le 0.3$) are able to produce a substantial reorientation.
In fact there is a class of active galaxies which exhibit radio
lobes at two, nearly-orthogonal orientations, and in which the
production of plasma along the fainter lobes appears to have ceased 
\cite{leahy-92,dennett-02}.
These ``X-shaped'' or ``winged'' radio galaxies, of which
about a dozen are known, are plausible sites of recent (within
the last $\sim 10^8$ yr) black-hole coalescence \cite{ekers-02,zier-02}.
Furthermore the implied coalescence rate is roughly consistent
with the expected merger rate for the host galaxies of luminous
radio sources \cite{ekers-02}.
Alternative models have been proposed for the X-shaped sources,
including a warping instability of accretion disks \cite{pringle-96},
backflow of gas along the active lobes \cite{leahy-84},
and binary-disk interactions before coalescence \cite{liu-04}.
\footnote{Liu (2004) criticized the black hole coalescence model
on the grounds that ``calculations based on
general relativity show that the change in inclination of a
rotating central SMBH is negligible in a minor merger and
a significant reorientation of the active SMBH requires a
comparatively rare major merger (Hughes \& Blandford 2003).''
This erroneous statement probably had its origin in the final
sentence of the Hughes \& Blandford paper, which states that
``An abrupt change in inclination...requires a comparatively
rare major merger.'' 
Hughes \& Blandford defined a ``major merger''
as having a mass ratio $q\ge 0.1$, in conflict with the standard 
definition among galactic dynamicists, $q\ge 0.3$.
In fact Hughes \& Blandford conclude, in agreement with Merritt
\& Ekers (2002), that mass ratios exceeding $\sim 0.2$ can result
in spin-flips.}
It is likely that all of these mechanisms are active at some level
and that
the time scale for realignment influences the radio source
morphology, with the most rapid realignments producing the
classical X-shaped sources, while slower realignment would
cause the jet to deposit its energy into a large volume, leading
to an S-shaped FRI radio source \cite{ekers-02}.

If the black holes are spinning prior to coalescence,
they will experience spin-orbit precession, on a time
scale that is intermediate between $t_{gr}$ and the 
orbital period.
To PN2.5 order, the spin angular momentum of either hole
evolves as \cite{kidder-95}
\beq
{d{\bf S}_1\over dt} = {1\over a^3}\left[\left(2+{3m_2\over 2m_1}\right) 
{\bf L}_{\rm orb} - {\bf S}_2 + 3\left(\hat n\cdot {\bf S}_2\right)\hat n\right]\times {\bf S}_1
\eeq
where $\hat n$ is a unit vector in the direction of the displacement
vector between the two black holes.
The evolution equation for ${\bf S}_2$ is given by interchanging
the indices.
The magnitude of each spin vector remains fixed 
(to this order), and each spin precesses around the total angular
momentum vector ${\bf J}={\bf L}_{\rm orb}+{\bf S}_1 + {\bf S}_2$.
When the two black holes are comparably massive,
the orbital angular momentum greatly exceeds the spin angular
momentum of either hole until just prior to coalescence.
As the binary shrinks, the spins have a tendency to unalign
with ${\bf L}_{\rm orb}$.
Ignoring spin-spin effects, the precession rate for equal-mass
holes in a circular orbit is
\beq
\Omega_1 = {7\over 2}{m_{12}^{1/2}\mu\over a^{5/2}}.
\eeq
The precession rate is lower than the orbital frequency:
\beq
{\Omega_{\rm orb}\over\Omega_1} = 
{2\over 7}{a\over \mu} = {16\over 7}{a\over R}
\eeq
where $R=2m_{12}$,
but higher than the radiation reaction time scale:
\beq
{1\over 2\pi}\Omega_1t_{gr} = {35\over 1024\pi}
{a^{3/2}\over m_{12}^{3/2}} = {35\sqrt{2}\over 512\pi}
\left({a\over R}\right)^{3/2},
\eeq
unless the binary is close to coalescence.
If the dominant source of energy loss during the late stages
of infall is gravitational radiation, the spin direction will
undergo many cycles of precession before the black holes
coalesce.
This does not seem to happen in the $X$-shaped radio sources,
based on the apparently sudden change in jet direction;
if the coalescence model for the $X$-sources is correct, the final stages
of infall must occur on a shorter time scale than $t_{gr}$.
This might be seen as evidence that shrinkage of the binary is usually driven
by gas dynamics, not gravitational radiation losses, prior to the final coalescence.

When the black hole masses are very different,
$q\ll 1$, the ratio of spin of the larger hole to ${\bf L}_{\rm orb}$
is
\beq
{S_1\over L_{\rm orb}} = {f_1\over q} \left({m_1\over a}\right)^{1/2} =
{f_1\over q}\left({R\over a}\right)^{1/2}
\eeq
where $S_1=f_1m_1^2$.
The two quantities are approximately equal when the separation
measured in units of the larger hole's Schwarzschild radius
is equal to $q^{-2}$.
When this separation is reached, the binary orbit rapidly changes 
its plane, and a new
regime is reached where the spin of the smaller black hole
precesses about the spin of the larger hole.
The precession rate of the larger hole is given by
\beq
\Omega_1 = {2m_1^{1/2}m_2\over a^{5/2}}
\eeq
in both regimes, and
\beq
{\Omega_{\rm orb}\over \Omega_1} = {2a\over R},\ \ \ \ 
{1\over 2\pi}\Omega_1t_{gr} = 
{5\over 128\pi}\left({a\over R}\right)^{3/2}.
\eeq

The spin direction of a black hole formed via binary coalescence
is also affected by torques that reorient the binary prior to
coalescence.
The role of torques from gaseous accretion disks 
was discussed above; another source of torques
is perturbations from passing stars or gas clouds \cite{merritt-02}.
A single star that passes within a distance $\sim 3a$
of the binary will exchange orbital angular momentum
with it, leading both to a change in the binary's 
orbital eccentricity as well as a change in the
orientation of the binary's spin axis, as discussed above.
Referring to equations (\ref{eq:def_H}) and (\ref{eq:def_L}),
the reorientation rate is related to the hardening rate
via
\beq
{\langle\Delta\vartheta^2\rangle}= {L\over H}{\ms\over\m12}t^{-1}_{\rm harden}
\eeq
where $t^{-1}_{\rm harden}=a(d/dt)(1/a)$.
Scattering experiments \cite{merritt-02} give $L/H\approx 4$
for a hard, equal-mass binary.
The implied change in the binary's orientation after shrinking
from $a\approx a_h$ to $a\approx 10^{-3}a_h$ is
\beq
\delta\theta \approx \sqrt{30\ms\over\m12}.
\eeq
The reorientation begins to be significant if $\ms/\m12\gap 10^{-3}$,
which may be the case for intermediate-mass black holes.

\section{Summary}
\begin{enumerate}
\item No completely convincing case of a bound, binary SBH
has yet been disccovered.
``Dual'' SBHs, i.e. two, widely-separated SBHs in a single system,
have been seen in some interacting galaxies and binary quasars.
However none has a projected separation less than $\sim 1$ kpc.
Strong limits can be placed on the binarity of the Milky Way SBH.
\item The evolution of binary SBHs in gas-poor galaxies is dominated
by the gravitational slingshot ejection of stars that pass near
the binary, carrying away energy and angular momentum.
Once the binary has ejected all stars on intersecting orbits,
continued hardening depends on a refilling of the binary's
``loss cone.'' 
Possible mechanisms for loss-cone refilling include
star-star gravitational scattering, chaotic orbits in
non-axisymmetric potentials, and perturbations from
additional massive objects.
\item Most $N$-body simulations of binary evolution have been
based on such small particle numbers that the binary's loss
cone was refilled at a spuriously high rate by gravitational
encounters, Brownian motion of the binary and other finite-$N$
effects.
As a result, many results from these studies can not 
usefully be extrapolated to
the large-$N$ regime; in particular, they do not make useful
predictions about the binary hardening rates expected in real
galaxies.
\item $N$-body studies do appear to make robust predictions
about the ``mass deficit,''  the mass in stars ejected
from a galactic nucleus by an evolving binary.
Observed mass deficits are of order $1-2$ times the SBH mass,
consistent with $N$-body predictions.
\item Studies of the interaction of binary SBHs with gas
are in their infancy.
Major uncertainties are the amount, distribution and thermodynamic
state of gas very near the centers of galaxies containing massive
binaries.
\item Binary coalescence can have a large influence on SBH spins;
even mass ratios as extreme as $10:1$ can substantially
spin up or re-orient a SBH.
Evidence for spin-flips may have been observed in the so-called
$X$-shaped radio sources.

\end{enumerate}

\section{Acknowledgements}
\label{section:acknowledgements}
This work was supported by grants 
AST-0071099, AST-0206031, AST-0420920 and AST-0437519 from the 
NSF, grant 
NNG04GJ48G from NASA,
and grant HST-AR-09519.01-A from
STScI.  M.~M.~was supported at Caltech by a postdoctoral fellowship from the Sherman Fairchild Foundation.

\newpage

%===================================================================================

% To use the plain latex bibliography in 'livrev_template_plain.bbl' do:
% 'cp LivRevTemplate_plain.bbl LivRevTemplate.bbl'
% 'latex LivRevTemplate'
% 'latex LivRevTemplate'
% 'latex LivRevTemplate'

% To use the bibtex bibliography in 'LivRevTemplate.bib' do:
% 'latex LivRevTemplate'
% 'bibtex LivRevTemplate'
% 'latex LivRevTemplate'
% 'latex LivRevTemplate'

\bibliography{astroph}

\end{document}